\documentclass[showpacs,preprintnumbers,amsmath,amssymb]{revtex4}

\usepackage{graphicx}
\usepackage{dcolumn}
\usepackage{bm}
\begin{document}
\title{Theoretical study of lepton events in the atmospheric neutrino experiments at SuperK}
\author{M. Sajjad Athar, S. Chauhan and S. K. Singh}
\affiliation{Department of Physics, Aligarh Muslim University, Aligarh-202 002, India}
\email{sajathar@rediffmail.com}
\date{\today}
\begin{abstract}
Super-Kamiokande has reported the results for the lepton events in the atmospheric neutrino experiment. These results have been presented for a 22.5kT water fiducial mass on an exposure of 1489 days, and the events are divided into sub-GeV, multi-GeV and PC events. We present a study of nuclear medium effects in the sub-GeV energy region of atmospheric neutrino events for the quasielastic scattering, incoherent and coherent pion production processes, as they give the most dominant contribution to the lepton events in this energy region. We have used the atmospheric neutrino flux given by Honda et al. These calculations have been done in the local density approximation. We take into account the effect of Pauli blocking, Fermi motion, Coulomb effect, renormalization of weak transition strengths in the nuclear medium in the case of the quasielastic reactions. The inelastic reactions leading to production of leptons along with pions is calculated in a $\Delta $- dominance model by taking into account the renormalization of $\Delta$ properties in the nuclear medium and the final state interaction effects of the outgoing pions with the residual nucleus. We present the results for the lepton events obtained in our model with  and without nuclear medium effects, and compare them with the Monte Carlo predictions used in the simulation and the experimentally observed events reported by the Super-Kamiokande collaboration.
\end{abstract}
\pacs{12.15.-y,13.15+g,13.60Rj,23.40.Bw,25.30Pt}
\maketitle
\section{Introduction}
There are now many evidences that neutrinos oscillate and the neutrinos are not massless. These  come from the experiments performed with atmospheric~\cite{Atmos1}-\cite{Atmos7}, reactor~\cite{Reac1},\cite{Reac2}, solar~\cite{solar1}, \cite{solar2} neutrinos and neutrinos obtained from the accelerators as in the experiments performed by MiniBooNE~\cite{mini}-\cite{minicite}, SciBooNE~\cite{Sciboonecite}, K2K~\cite{Ahn}-\cite{k2kcite} and MINOS collaboration~\cite{minos1}-\cite{minoscite}. The information from these experiments puts the limits on the solar and atmospheric neutrino mass differences viz. ${{\Delta m}^2_{solar}}= 7.65^{+0.23}_{-0.20}\times 10^{-5} eV^2$ and $|{{\Delta m}_{atm}}|^2= 2.4 ^{+0.12}_{-0.13} \times 10^{-3} eV^2$. Two out of the four parameters of the Pontecorvo-Maki-Nakagawa-Sakata (PMNS) matrix i.e. $sin^2(\theta_{12})=0.304^{+0.022}_{-0.016}$ and $sin^2(2\theta_{23})=0.50^{+0.07}_{-0.06}$ \cite{neutpara} are known with a better accuracy while $sin^2(\theta_{13})$ and CP violating phase $\delta$ are still unknown. There is some information on $sin^2(\theta_{13})$ ($<$ 0.01 at 1$\sigma$), but there is no information on $\delta$. With these experimental observations lot of theoretical as well as experimental activities are going on. Now the aim of the experimentalists is to determine with better precision the various parameters of the PMNS matrix, absolute masses of the different flavors of neutrinos, to see whether the neutrino mass hierarchy is normal or inverted and the CP violation in the neutrino sector exists or not. On the theoretical side recently it has been emphasized in a series of neutrino workshops and conferences like NuInt~\cite{nuint}, NuFact~\cite{nufact}, NOW~\cite{now}, etc. that at the neutrino energies of a few GeV energy region the study of neutrino nucleus cross section is very important which is relevant for the experiments with atmospheric neutrinos at Super-Kamiokande\cite{Atmos2}, ICARUS\cite{ICARUS}, accelerator neutrinos at MiniBooNE, K2K, T2K~\cite{t2k}-\cite{t2kcite}, NOvA~\cite{nova} and the future experiments planned with Beta beams~\cite{betabeams1}-\cite{Bando} and superbeams at Neutrino Factories~\cite{Albright},\cite{Bando}.  Neutrino-Nucleus cross section is one of the inputs in predicting the neutrino event rates. In the neutrino nuclear scattering process nuclear medium effects should be taken into account for writing the neutrino generator Monte Carlo codes which are used in analyzing the neutrino oscillation experiments. These oscillation experiments use various nuclear targets like $^{12}C$, $^{16}O$, $^{40}Ar$, $^{56}Fe$, etc. For example, Super-Kamiokande(SuperK) is a 50kT water Cerenkov detector observing neutrinos from the terrestrial as well as accelerator neutrino sources. The nuclear target is $^{16}O$ in water($H_2O$), and neutrinos(antineutrinos) are interacting with the free protons as well as with the nucleons inside the oxygen nucleus. The Monte Carlo simulation of  the lepton events uses Smith-Moniz model \cite{SmithMoniz} for the quasielastic process which does not include the effect of nuclear medium arising due to nucleon-nucleon correlations but includes only the effect of Pauli principle and Fermi motion in a Fermi gas model. In the case of inelastic reaction for incoherent and coherent pion production it uses Rein and Sehgal model~\cite{Rein, Rein1} with the inclusion of the nuclear effects arising due to final state interactions of pions with the nucleus like pion absorption and pion scattering. Furthermore, there are also some recent calculations on multipion production and deep inelastic neutrino reactions showing that the nuclear effects can be important in these energy regions.

In this work, we have studied the lepton event rates for the atmospheric neutrinos at SuperK with and without the nuclear medium effects and compared our results with the experimental observed events and also with the Monte Carlo predictions for the events used by SuperK collaboration~\cite{Atmos2}. These results have been presented for a 22.5kT water fiducial mass on an exposure of 1489 live days, and we have taken the sub-GeV events in our analysis which have been classified as the region in which lepton's energy $E_l < 1.33GeV$ and minimum observed momenta of electrons and muons are 100MeV and 200MeV respectively~\cite{Atmos2}. Our Monte-Carlo analysis of the events has been done by considering the nuclear medium effects in the quasielastic, incoherent and coherent pion production processes, as they give the most dominant contribution in the sub-GeV energy region of atmospheric neutrino events. We have used the atmospheric neutrino flux given by Honda et al.~\cite{Honda1},\cite{Honda2}. 

In the case of quasielastic reaction, the effects of Pauli principle and Fermi motion are included through the Lindhard function calculated in a local density approximation. The renormalization of the weak transition strengths is calculated in the Random Phase Approximation(RPA) through the interaction of the p-h excitations as they propagate in the nuclear medium using a nucleon-nucleon potential described by pion and rho exchanges. The single pion production in the sub-GeV region is dominated by the resonance production in which a $\Delta$ resonance is excited and decays subsequently to a pion and a nucleon. When this process takes place inside the nucleus, there are two possibilities i.e. the target nucleus remains in the ground state leading to coherent production of pions or is excited and/or broken up leading to incoherent production of pions. We have considered both the production processes in the $\Delta$ resonance model in the local density approximation to calculate single pion production accompanied by a lepton from the oxygen nucleus. The effect of nuclear medium on the production of $\Delta$ is treated by including the modification of $\Delta$ properties in the medium. Once pions are produced, they undergo final state interactions with the residual nucleus. We have taken the final state interaction effects for both the incoherent and coherent pion production processes. This work is based on our study of the nuclear medium effects in the neutrino(antineutrino) induced reaction on the various nuclear targets like $^{12}C$, $^{16}O$, $^{40}Ar$, $^{56}Fe$ etc., in the local density approximation, which have been applied to low and intermediate energy neutrinos for the charged current quasielastic process and the inelastic pion production process for the incoherent and coherent lepton production accompanied by a pion~\cite{Quasi1}-\cite{Coh1}.

The plan of presentation is as follows. In section-2, we describe the charged current neutrino(antineutrino) induced quasielastic inclusive production of leptons from the nucleus. In section-3, we describe the charged current neutrino(antineutrino) induced inelastic production of leptons accompanied by a pion from the nucleus, where we describe the incoherent pion production as well as coherent pion production processes. In section-4, we present our results for the total scattering cross section as well as $Q^2$ distribution averaged over the atmospheric neutrino flux given by Honda et al.~\cite{Honda1},\cite{Honda2} for the SuperK site. Furthermore, we discuss the dependence of the cross sections and $Q^2$ distribution on the axial dipole mass $M_A$ and the various parameterization of the form factors discussed in the literature recently for the isovector form factors in the case of quasielastic scattering and N-$\Delta$ transition form factors in the case of inelastic scattering. We have averaged the total scattering cross section $\sigma(E)$ and $Q^2$-distribution over the atmospheric neutrino flux given by Honda et al.~\cite{Honda2} for the SuperK site to obtain the total lepton production event rate and the results have been compared with the observed numbers at SuperK and also with the numbers used by them in their Monte Carlo~\cite{Atmos2}. In section-5, we conclude our findings.
\section{{\bf QUASIELASTIC REACTION}}
The basic reaction for the quasielastic process is a neutrino interacting with a neutron inside the nucleus is given by
\begin{eqnarray}\label{quasi_reaction}
\nu_l(k) + n(p) \rightarrow l^{-}(k^\prime) + p(p^{\prime}); l=e^-,\mu^- 
\end{eqnarray}
The invariant matrix element for the charged current reaction of neutrino, given by Eq.(\ref{quasi_reaction}) is written as
\begin{eqnarray}\label{qe_lep_matrix}
{\cal M}=\frac{G_F}{\sqrt{2}}\cos\theta_C~l_\mu~J^\mu
\end{eqnarray}
where $G_F$ is the Fermi coupling constant (=1.16639$\times 10^{-5}$GeV$^{-2}$), $\theta_C(=13.1^0)$ is the Cabibbo angle and the leptonic weak current is given by
\begin{eqnarray}\label{lep_curr}
l_\mu&=&\bar{u}(k^\prime)\gamma_\mu(1 - \gamma_5)u(k)
\end{eqnarray}

$J^\mu$ is the hadronic current given by
\begin{equation}\label{had_curr}
J_\mu=\bar{u}(p')\left[F_1^V(q^2)\gamma_\mu+F_2^V(q^2)i\sigma_{\mu\nu}\frac{q^\nu}{2M} + F_A^V(q^2)\gamma_\mu\gamma_5 + F_P^V(q^2)q_\mu\gamma_5\right]u(p)
\end{equation}
where, $q^2=(k-k^\prime)^2$ is the momentum transfer square and $M$ is the nucleon mass. $F_{1,2}^V(q^2)$ are the isovector vector form factors and $F_A(q^2)$, $F_P(q^2)$ are respectively the isovector axial vector and pseudoscalar form factors. 

Using the leptonic and hadronic currents given in Eq.(\ref{lep_curr}) and Eq.(\ref{had_curr}), the matrix element square is obtained by using Eq.(\ref{qe_lep_matrix}):
\begin{equation}\label{mat_quasi}
{|{\cal M}|^2}=\frac{G_F^2}{2}\cos^2\theta_C~{ L}_{\mu\nu}^{(\nu)} {J}^{\mu\nu}
\end{equation} 
${ L}_{\mu\nu}^{(\nu)}$ is the leptonic tensor calculated to be
\begin{eqnarray}\label{lep_tens}
{L}_{\mu\nu}^{(\nu)}&=&{\bar\Sigma}\Sigma{l_\mu}^\dagger l_\nu=L_{\mu\nu}^{S~(\nu)} + i L_{\mu\nu}^{A~(\nu)},~~~~\mbox{where}\\
L_{\mu\nu}^{S~(\nu)}&=&8~\left[k_\mu k_\nu^\prime+k_\mu^\prime k_\nu-g_{\mu\nu}~k\cdot k^\prime\right]~~~~\mbox{and}\nonumber\\
L_{\mu\nu}^{A~(\nu)}&=&8~\epsilon_{\mu\nu\alpha\beta}~k^{\prime \alpha} k^{\beta}
\end{eqnarray}
For antineutrino induced reaction ${\bar\nu}_l(k) + p(p) \rightarrow l^{+}(k^\prime) + n(p^{\prime}); l=e^+,\mu^+ $, ${ L}_{\mu\nu}^{(\bar\nu)}$ is given by 
\[{ L}_{\mu\nu}^{(\bar\nu)} = { L}_{\nu\mu}^{(\nu)}\]

The hadronic tensor ${J}^{\mu\nu}$ is given by:
\begin{eqnarray}\label{had_tens}
J^{\mu\nu}&=&\bar{\Sigma}\Sigma J^{\mu\dagger} J^\nu\nonumber\\
&=&\frac{1}{2}\mbox{Tr}\left[({\not p^\prime}+M)\Gamma^\mu ({\not p}+M)\tilde\Gamma^\nu\right]
\end{eqnarray}
where
\begin{equation}\label{had_inter}
\Gamma_\mu=\left[F_1^V(q^2)\gamma_\mu+F_2^V(q^2)i\sigma_{\mu\nu}\frac{q^\nu}{2M} + F_A^V(q^2)\gamma_\mu\gamma_5 + F_P^V(q^2)q_\mu\gamma_5\right]\end{equation}
and $\tilde\Gamma^\nu=\gamma^0~{\Gamma^\nu}^\dagger~\gamma^0$\\
The hadronic current contains two isovector vector form factors $F_{1,2}^V(q^2)$ of the nucleons, which are given as
\begin{equation}\label{f1v_f2v}
F_{1,2}^V(q^2)=F_{1,2}^p(q^2)- F_{1,2}^n(q^2) 
\end{equation}
where $F_{1}^{p(n)}(q^2)$ and $F_{2}^{p(n)}(q^2)$ are the Dirac and Pauli form factors of proton(neutron) which are in turn expressed in terms of the experimentally determined Sach's electric $G_E^{p,n}(q^2)$ and magnetic $G_M^{p,n}(q^2)$ form factors of the nucleons given by
\begin{eqnarray}\label{gm_ge}
G_M^{p,n}(q^2)&=&F_1^{p,n}(q^2)-F_2^{p,n}(q^2)\\
G_E^{p,n}(q^2)&=&F_1^{p,n}(q^2)+\frac{q^2}{4M^2}~F_2^{p,n}(q^2)
\end{eqnarray} 
This results in the following form of the isovector vector form factors $F_{1,2}^V(q^2)$ to be used in Eq.(\ref{had_curr}) 
\begin{eqnarray}\label{f1pn_f2pn}
F_1^{p,n}(q^2)&=&\left(1-\frac{q^2}{4M^2}\right)^{-1}~\left[G_E^{p,n}(q^2)-\frac{q^2}{4M^2}~G_M^{p,n}(q^2)\right]\\
F_2^{p,n}(q^2)&=&\left(1-\frac{q^2}{4M^2}\right)^{-1}~\left[G_M^{p,n}(q^2)-G_E^{p,n}(q^2)\right]
\end{eqnarray}
$G_E^{p,n}(q^2)$ and $G_M^{p,n}(q^2)$ are described by Galster parameterization~\cite{Galster} with $Q^2=-q^2$
\begin{eqnarray}\label{gep_gmp}
G_E^{p}(Q^2)&=&G_D(Q^2),~~G_E^{n}(Q^2)=-\tau\mu_{n}G_D(Q^2)\xi_{n},~~~G_D(Q^2)=\left[1+\frac{Q^2}{M_V^2}\right]^{-2},\\
\xi_{n}&=&\frac{1}{1+\lambda_{n}\frac{Q^2}{4M^2}},~~ \lambda_{n}=5.6,~~\tau=\frac{Q^2}{4M^2}\\
\frac{G_M^{p}(Q^2)}{\mu_p}&=&G_D(Q^2),~~~~\frac{G_M^{n}(Q^2)}{\mu_n}=G_D(Q^2)
\end{eqnarray}
with proton and neutron magnetic moments as $\mu_p$=2.79285$\mu_N$ and $\mu_n$=-1.913$\mu_N$, respectively, $Q^2=-q^2$ and $M_V$=0.84GeV. 

The isovector axial form factor is parametrized as
\begin{equation}\label{fa}
F_A(Q^2)=F_A(0)~\left[1+\frac{Q^2}{M_A^2}\right]^{-2}
\end{equation}
and is obtained from the quasielastic neutrino and antineutrino scattering as well as from pion electroproduction data. We have used axial charge $F_A(0)$=-1.267 and the axial dipole mass $M_A$=1.1GeV, which is presently being used in the SuperK analysis~\cite{Mitsuka}. The world average value for $M_A$=1.026$\pm$0.020 GeV~\cite{wor_avg}, which is consistent with the recent NOMAD results for $M_A$=1.05$\pm$0.02$\pm$0.06 GeV~\cite{nomad} obtained from the quasielastic $\nu_\mu$ and ${\bar\nu}_\mu$ reactions with carbon, however, the values for axial dipole mass obtained by the experiments performed at K2K and MiniBooNE differ from this value. The values reported by the K2K experiments are $M_A$=1.14$\pm$0.11GeV\cite{k2k_scifi_scibar}~ from the SciBar detector and $M_A$=1.20$\pm$0.12GeV~\cite{k2k_scifi_scibar}  by the SciFi detector experiments and $M_A$=1.23$\pm$0.2GeV recently reported by the MiniBooNE collaboration~\cite{mini}.

The pseudoscalar form factor $F_p^V(Q^2)$ is dominated by the pion pole and is given in terms of $F_A^V(Q^2)$ using the Goldberger-Treiman relation as 
\begin{equation}\label{fp}
F_p^V(Q^2)=\frac{2MF_A^V(Q^2)}{m_\pi^2+Q^2}
\end{equation}
Recently several new parameterizations for electromagnetic isovector form factors ~\cite{BBA03}-\cite{Alberico} have been presented which are obtained from the fits to the electron scattering data. To see the dependence of the cross section on the various parameterizations of the electromagnetic form factors we have used the parameterizations given by Budd et al.~\cite{BBA03} known as BBA-03, Bradford et al.~\cite{BBBA05} known as BBBA-05, Bosted~\cite{Bosted} as well as the parameterization given by Alberico et al.~\cite{Alberico} 

The form of the electric and magnetic Sach's form factor given by Budd et al.~\cite{BBA03} (BBA03) for the nucleon is 
\begin{eqnarray}\label{ge_gm_bba01}
G_{E}^{p}(Q^2)&=&\frac{1}{1+3.253Q^2+1.422Q^4+0.08582Q^6+0.3318Q^8-0.09371Q^{10}+0.01076Q^{12}}\\ \nonumber
\frac{G_{M}^{p}(Q^2)}{\mu_{p}}&=&\frac{1}{1+3.104Q^2+1.428Q^4+0.1112Q^6-0.006981Q^8+0.0003705Q^{10}-0.7063E{-05}Q^{12}}\\ \nonumber
\frac{G_{M}^{n}(Q^2)}{\mu_{n}}&=&\frac{1}{1+3.043Q^2+0.8548Q^4+0.6806Q^6-0.1287Q^8+0.008912Q^{10}}\\ \nonumber
\frac{G_{E}^{n}(Q^2)}{\mu_{n}}&=&-\frac{0.942\tau}{1+4.61\tau}G_D(Q^2) 
\end{eqnarray}
The form of electric and magnetic Sach's form factor given by Bradford et al.~\cite{BBBA05} (BBBA-05) is
\begin{eqnarray}\label{ge_gm_bbba01}
G_{E}^{p}(Q^2)&=&\frac{1-0.0578\tau}{1+11.1\tau+13.6\tau^2+33.0\tau^3}\\ \nonumber
\frac{G_{M}^{p}(Q^2)}{\mu_{p}}&=&\frac{1+0.150\tau}{1+11.1\tau+19.6\tau^2+7.54\tau^3}\\ \nonumber
G_{E}^{n}(Q^2)&=&\frac{1.25\tau+1.30\tau^2}{1-9.86\tau+305\tau^2-758\tau^3+802\tau^4}\\ \nonumber
\frac{G_{M}^{n}(Q^2)}{\mu_{n}}&=&\frac{1+1.81\tau}{1+14.1\tau+20.7\tau^2+68.7\tau^3}
\end{eqnarray}
The parametrization given by Bosted~\cite{Bosted} is 
\begin{eqnarray}\label{gep_bosted}
G_E^p(Q^2)&=&\frac{1}{1+0.62Q+0.68Q^2+2.80Q^3+0.83Q^4}\\ \nonumber
\frac{G_M^p(Q^2)}{\mu_p}&=&\frac{1}{1+0.35Q+2.44Q^2+0.50Q^3+1.04Q^4+0.34Q^5}\\\nonumber
\frac{G_M^n(Q^2)}{\mu_n}&=&\frac{1}{1-1.74Q+9.29Q^2-7.63Q^3+4.63Q^4}\\\nonumber
G_{E}^{n}(Q^2)&=&\frac{-1.25 \mu_{n}\tau G_{D}(Q^2)}{1+18.3\tau}
\end{eqnarray}
Alberico et al.~\cite{Alberico} have parameterized recently the electromagnetic form factors of the nucleon based on the recent experiments performed at Bates, MAMI and JLab.
The form of electric and magnetic Sach's form factor given by them~\cite{Alberico}  is
\begin{eqnarray}\label{ge_gm_alberico1}
G_{E}^{p}(Q^2)&=&\frac{1-0.14\tau}{1+11.18\tau+15.18\tau^2+23.57\tau^3}\\ \nonumber
\frac{G_{M}^{p}(Q^2)}{\mu_{p}}&=&\frac{1+1.07\tau}{1+12.30\tau+25.43\tau^2+30.39\tau^3}\\ \nonumber
\frac{G_{M}^{n}(Q^2)}{\mu_{n}}&=&\frac{1+2.13\tau}{1+14.53\tau+22.76\tau^2+78.29\tau^3}\\ \nonumber
G_{E}^{n}(Q^2)&=&\frac{-0.10}{(1+2.83Q^2)^2}+\frac{0.10}{(1+0.43Q^2)^2}
\end{eqnarray}
Using these parameterizations of the isovector form factors discussed above, we calculate the hadronic tensor given by Eq.(\ref{had_tens}) and the matrix element square using Eq.(\ref{mat_quasi}). With this ${|{\cal M}|^2}$ the charged current quasielastic lepton production  cross section is calculated. 

The general expression for the differential cross section for the reaction shown in Eq.(\ref{quasi_reaction}) is given by
\begin{eqnarray}\label{diff_xsect_quasi}
d\sigma=\left(\frac{G_F^2\cos^2\theta_C}{2}\right)\frac{(2\pi)^{4}\delta^{4}(k+p-p^\prime-k^\prime)}{4\sqrt{(k\cdot k^\prime)^{2}-m_{\nu}^{2}M_n^{2}}}\frac {d^{3}{\bf{k^\prime}}}{(2\pi)^{3}2E_{l}}\frac {d^{3}{\bf p^{\prime}}}{(2\pi)^{3}2E_{p}}\prod_f(2m_f){ L}_{\mu\nu} {J}^{\mu\nu}
\end{eqnarray}
where f is the number of fermions in the final state.
The double differential cross section  $\sigma_0(E_e,|\vec k^\prime|)$ for the basic reaction is then written as
\begin{equation}\label{sig_zero}
\sigma_0(E_l,|\vec k^\prime|)=\frac{{|\vec k^\prime|}^2}{4\pi E_{\nu_l} E_l}\frac{M_n M_p}{E_n E_p}{\bar\Sigma}\Sigma{|{\cal M}|^2}\delta[q_0+E_n-E_p]
\end{equation}
In a nucleus, the neutrino scatters from a neutron moving in the finite nucleus of neutron density $\rho_n(r)$, with a local occupation number $n_n({\bf{p}},{\bf{r}})$. In the local density approximation the scattering cross section is written as
\begin{equation}\label{sig_4}
\sigma(E_l,|\vec k^\prime|)=\int 2d{\bf r}d{\bf p}\frac{1}{(2\pi)^3}n_n({\bf p},{\bf r})\sigma_0(E_l,k^\prime)
\end{equation}
where $\sigma_0(E_l,|\vec k^\prime|)$ is given by Eq.(\ref{sig_zero}). The neutron energy $E_n$ and proton energy $E_p$ are replaced by $E_n(|\vec p|)$ and $E_p(|\vec{p}+\vec{q}|)$, where $\bf{p}$ is now the momentum of the target neutron inside the nucleus. Inside the nucleus the neutrons and protons are not free and their momenta are constrained to satisfy the Pauli principle, i.e., ${p_{n}}<{p_{F_{n}}}$ and ${p^\prime}_{p}(=|{\bf p}_{n}+{\bf q}|) > p_{F_{p}}$, where $p_{F_n}$ and $p_{F_p}$ are the local Fermi momenta of neutrons and protons at the interaction point in the nucleus and are given by $p_{F_n}=\left[3\pi^2\rho_n(r)\right]^\frac{1}{3}$ and $p_{F_p}=\left[3\pi^2\rho_p(r)\right]^\frac{1}{3}$, $\rho_n(r)$ and $\rho_p(r)$ are the neutron and proton nuclear densities which are given in terms of the nuclear density of the oxygen nucleus
\begin{equation}\label{rho}
\rho_{n}(r)=\frac{(A-Z)}{A}\rho(r);~~~\rho_{p}(r)=\frac{Z}{A}\rho(r)
\end{equation}
where $\rho(r)$ is the density of oxygen nucleus taken to be 3 parameter Fermi(3pF) density and the density parameters have been taken from Ref.~\cite{Vries}. There are also other parametrizations for the nuclear ($^{16}O$) density available in the literature like Harmonic Oscillator (HO) density~\cite{Vries} and Modified Harmonic Oscillator (MHO)~\cite{Vries} density and we have also studied the dependence of the cross section on the various nuclear densities. Furthermore, in nuclei the threshold value of the reaction i.e. the Q-value of the reaction($Q_{r}$) has to be taken into account, which we have taken to be the value corresponding to the lowest allowed Fermi transition.

\begin{center}
\begin{figure}
\includegraphics{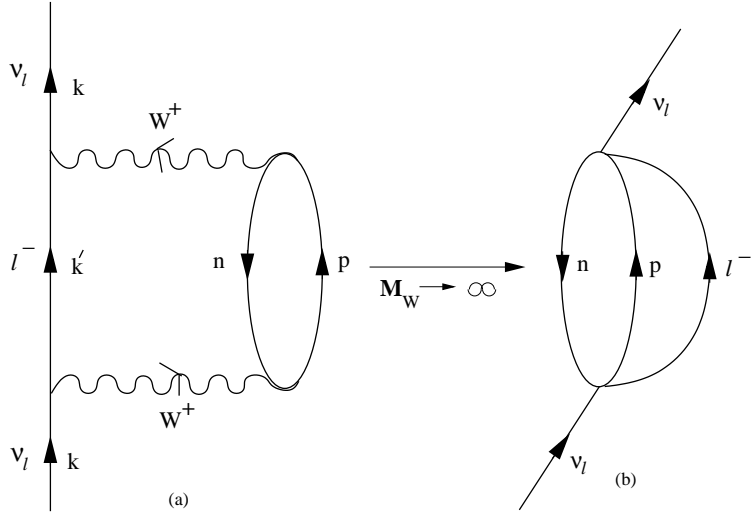}
\caption{Diagrammatic representation of the neutrino self-energy diagram corresponding to the ph-excitation leading to $\nu_l +n \rightarrow l^- + p$ in nuclei. In the large mass limit of the IVB(i.e.$M_W\rightarrow \infty$) the diagram 1(a) is reduced to 1(b) which is used to calculate ${|{\cal M}|^2}$ in Eq.(\ref{mat_quasi}).}
\label{fg:fig1}
\end{figure}
\end{center}

These considerations lead to a modification in the $\delta$ function used in Eq.(\ref{sig_zero}) i.e.  $\delta[q_0+E_n-E_p]$ is modified to $\delta[q_0+E_n(\vec{p})-E_p(\vec{p}+\vec{q})-Q_{r}]$ and the factor
\begin{equation}\label{delta_modi}
\int \frac{d\bf{p}}{(2\pi)^3}{n_n(\bf{p},\bf{r})}\frac{M_n M_p}{E_n E_p}\delta[q_0+E_n-E_p]
\end{equation}
occurring in Eq.(\ref{sig_4}) is replaced by $-(1/{\pi})$Im${{U_N}(q_0,\vec{q})}$, where ${{U_N}(q_0,\vec{q})}$ is the Lindhard function corresponding to the particle hole(ph) excitation\cite{Quasi4} shown in Fig.(\ref{fg:fig1}) and is given by
\begin{equation}\label{lindhard}
{U_N}(q_0,\vec{q}) = {\int \frac{d\bf{p}}{(2\pi)^3}\frac{M_nM_p}{E_nE_p}\frac{n_n(p)\left[1-n_p(\vec p + \vec q) \right]}{q_0+{E_n(p)}-{E_p(\vec p+\vec q)}+i\epsilon}}
\end{equation}
where $q_0$=$E_{\nu_l}-E_l-Q_{r}$. For the antineutrino reaction the suffix n and p will get interchanged.
The imaginary part of the Lindhard function is obtained to be~\cite{Quasi4}: 
\begin{equation}\label{lindhard_imag}
Im{U_N}(q_0, \vec{q}) = -\frac{1}{2\pi}\frac{M_p{M_n}}{|\vec{q}|}\left[E_{F_1}-A\right]
\end{equation}
with $q^2<0$, $E_{F_2}-q_0<E_{F_1}$ and $\frac{-q_0+|\vec{q}|{\sqrt{1-\frac{4{M^2}}{q^2}}}}{2}<{E_{F_1}}$, 
where $E_{F_1}=\sqrt{p{_{F_n}}^2+{M_n}^2}$, $E_{F_2}=\sqrt{{p_{F_p}}^2+{M_p}^2}$ and \\
A = $Max\left[M_n,\hspace{2mm}E_{F_2}-q_0,\hspace{2mm}\frac{-q_0+|\vec{q}|\sqrt{1-\frac{4{M^2}}{q^2}}}{2}\right]$.

\begin{figure}
\includegraphics{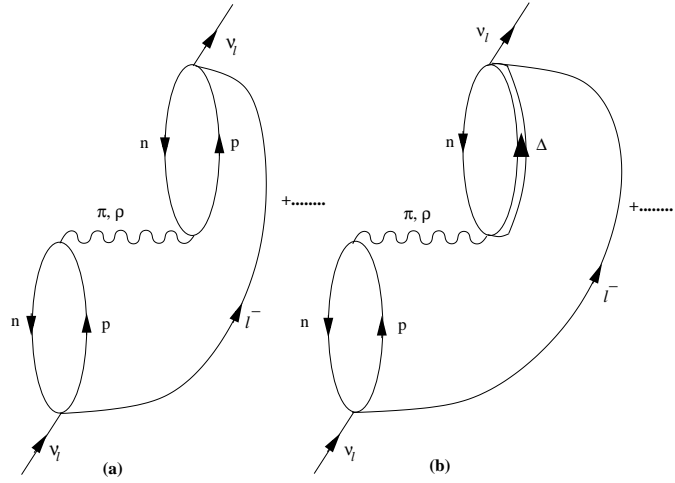}
\caption{ Many body Feynman diagrams (drawn in the limit $M_W\rightarrow \infty$) accounting for the medium polarization effects contributing to the process $\nu_l +n \rightarrow l^- + p$ transitions}
\label{fg:fig2}
\end{figure}

With inclusion of these nuclear effects the cross section $\sigma(E_\nu)$ is written as
\begin{eqnarray}\label{xsection_medeffects}
\sigma(E_\nu)&=&-2{G_F}^2\cos^2{\theta_c}\int^{r_{max}}_{r_{min}} r^2 dr \int^{{k^\prime}_{max}}_{{k^\prime}_{min}}k^\prime dk^\prime \int_{Q^{2}_{min}}^{Q^{2}_{max}}dQ^{2}\frac{1}{E_{\nu_l}^{2} E_l} L_{\mu\nu}J^{\mu\nu} Im{U_N}[E_{\nu_l} - E_l - Q_{r}, \vec{q}].
\end{eqnarray}
The outgoing lepton when comes out of the nucleus, its energy and momentum are modified due to the Coulomb interaction. The Coulomb distortion effect on the outgoing lepton has been taken into account in an effective momentum approximation(MEMA)~\cite{Quasi4},\cite{Engel} in which the lepton momentum and energy are modified. In the local density approximation, the effective energy of the lepton in the Coulomb field of the final nucleus is given by:
\[ E_{eff} = E_l + V_c(r), \]
where 
\begin{equation}\label{effective_coulomb}
V_c(r)=Z_f\alpha4\pi\left(\frac{1}{r}\int_0^r\frac{\rho_p(r^\prime)}{Z_f}{r^\prime}^2dr^\prime + \int_r^\infty\frac{\rho_p(r^\prime)}{Z_f}{r^\prime}dr^\prime\right)
\end{equation}
This leads to a change in the Imaginary part of the Lindhard function occurring in Eq.~(\ref{xsection_medeffects})
\begin{equation*}\label{changed_lindhard}
Im{U_N}(E_{\nu_l} - E_l - Q_{r}, {\bf q}) \rightarrow Im{U_N}(E_{\nu_l} - E_l - Q_{r} - V_c(r), {\bf q})
\end{equation*}
In the nucleus the strength of the electroweak coupling may change from their free nucleon values due to the presence of strongly interacting nucleons. Conservation of Vector Current (CVC) forbids any change in the charge coupling while magnetic and axial vector couplings are likely to change from their free nucleon values. These changes are calculated by considering the interaction of ph excitations in the nuclear medium in Random Phase Approximation (RPA) as shown in Fig.\ref{fg:fig2}. The diagram shown in Fig.\ref{fg:fig2} simulates the effects of the strongly interacting nuclear medium at the weak vertex. The ph-ph interaction is shown by the wavy line in Fig.\ref{fg:fig2} and is described by the $\pi$ and $\rho$ exchanges modulated by the effect of short range correlations.

The weak nucleon current described by Eq.(\ref{had_curr}) gives, in the non-relativistic limit, terms like $F_A \vec{\sigma}\tau_+$ and $i F_2 \frac{\vec{\sigma}\times \vec{q}}{2M}\tau_+$ which generate spin-isospin transitions in nuclei. While the term $i F_2 \frac{\vec{\sigma}\times \vec{q}}{2M}\tau_+$ couples to the transverse excitations, the term  $F_A \vec{\sigma}\tau_+$ couples to the transverse as well as longitudinal channels. These channels produce different RPA responses in the longitudinal and transverse channels when the diagrams of Fig.\ref{fg:fig2} are summed over. This is illustrated by considering the contribution of a term like $F_A\sigma^i$ in Eq.(\ref{had_curr}). The leading order contribution of this term to the hadronic tensor $J^{ij}$ in the medium is proportional to $F^2_A \delta_{ij}Im U_N$ which is now split between the longitudinal and transverse components as 
\begin{equation}\label{f2a_modify}
F^2_A\delta_{ij}Im {U_N}\rightarrow F^2_A \left[{\bf{\hat{q_i}}{\hat{q_j}}}+(\delta_{ij}-{\bf{\hat{q_i}}{\hat{q_j}}})\right]Im {U_N}
\end{equation}
The RPA response of this term after summing the higher order diagrams like Fig.\ref{fg:fig2} is modified and is given by $J^{ij}_{RPA}$
\begin{equation}\label{f2a_rpa}
J^{ij}\rightarrow J^{ij}_{RPA}= F^2_A{Im U_N}\left[\frac{{\bf{\hat{q_i}}{\hat{q_j}}}}{1-U_NV_l}+\frac{\delta_{ij}-{\bf{\hat{q_i}}{\hat{q_j}}}}{1-U_NV_t}\right]
\end{equation}
where $V_l$ and $V_t$ are the longitudinal and transverse parts of the nucleon-nucleon potential calculated with $\pi$ and $\rho$ exchanges and are given by 
\begin{eqnarray}\label{longi_part}
V_l(q) = \frac{f^2}{m_\pi^2}\left[\frac{q^2}{-q^2+m_\pi^2}{\left(\frac{\Lambda_\pi^2-m_\pi^2}{\Lambda_\pi^2-q^2}\right)^2}+g^\prime\right],\nonumber\\
V_t(q) = \frac{f^2}{m_\pi^2}\left[\frac{q^2}{-q^2+m^2_\rho}{C_\rho}{\left(\frac{{\Lambda_\rho}^2-m^2_\rho}{{\Lambda_\rho}^2-q^2}\right)^2}+g^\prime\right]\end{eqnarray}
$\Lambda_\pi=1.3 GeV$, $C_\rho=2$, $\Lambda_\rho=2.5GeV$, $m_\pi$ and $m_\rho$ are the pion and rho meson masses, and $g^\prime$ is the Landau-Migdal parameter taken to be $0.7$ which has been used quite successfully to explain many electromagnetic and weak processes in nuclei~\cite{Mukho},\cite{Gill}. Recently, in a work by Nieves et al.~\cite{Nieves2}, $g^\prime$ has been taken as 0.63. We have studied the dependence of cross section on the Landau-Migdal parameter by varying $g^\prime$ by 10$\%$ i.e. taking $g^{\prime}$ = 0.63 and 0.77.

This modified tensor $J^{ij}_{RPA}$ when contracted with the leptonic tensor $L_{ij}$ gives the contribution of the $F^2_A$ term to the RPA response. 
The effect of the $\Delta$ degrees of freedom in the nuclear medium is included in the calculation of the RPA response by considering the effect of ph-$\Delta$h and $\Delta$h-$\Delta$h excitations as shown in Fig.\ref{fg:fig2}(b). This is done by replacing $U_N$ by $U_N=U_N+U_\Delta$, where $U_\Delta$ is the Lindhard function for $\Delta$h excitation in the medium and the expressions for $U_N$ and $U_\Delta$ are taken from the Ref.~\cite{Oset1}. The different couplings of $N$ and $\Delta$ are incorporated in $U_N$ and $U_\Delta$ and then the same interaction strengths $V_l$ and $V_t$ are used to calculate the RPA response. These effects have been recently discussed by Nieves et al.~\cite{Nieves2}.

Thus, in the presence of nuclear medium effects, the total cross section $\sigma(E_\nu)$, is written as
 \begin{eqnarray}\label{cross_section_quasi}
\sigma(E_\nu)=-2{G_F}^2\cos^2{\theta_c}\int^{r_{max}}_{r_{min}} r^2 dr \int^{{k^\prime}_{max}}_{{k^\prime}_{min}}k^\prime dk^\prime \int_{Q_{min}^{2}}^{Q_{max}^{2}}dQ^2\frac{1}{E_{\nu_l}^2 E_l}L_{\mu\nu}{J^{\mu\nu}_{RPA}} Im{U_N}[E_{\nu_l} - E_l - Q_{r} - V_c(r), \vec{q}]
\end{eqnarray}
where $J^{\mu\nu}_{RPA}$ is the modified hadronic tensor when RPA effects are incorporated.
\section{INELASTIC PION PRODUCTION}
In the case of inelastic production process of leptons, the leptons are produced along with one or multiple pions. Around the energy region of 1 GeV, in the energy spectrum of atmospheric neutrino, the inelastic lepton production is dominated by the processes where lepton is accompanied by a single pion. These pion production processes take place mainly through the excitation of $\Delta$ which subsequently decay into a pion and a nucleon.     
When these processes take place inside the nucleus then either the target nucleus remains in the ground state where it does not change its identity leading to coherent production of pions or the target nucleus is excited and/or broken up leading to incoherent production of pions. Recently, two pion production and quasielastic hyperons~\cite{Hernandez, Singh} have been studied and are found to make small contribution to the lepton production in the energy region of present interest.

The basic reaction for the charged current neutrino(antineutrino) induced one pion production (CC1$\pi$) in nuclei, is that a neutrino(antineutrino) interacts with a nucleon N. The various possible channels contributing for one pion production processes are
\begin{eqnarray}\label{chan_numu_pi+}
\nu_l(k)+p(p)&\rightarrow& l^{-}(k^{\prime})+\Delta^{++}(P)\\
           &&~~~~~~~~~~~~~~                     \searrow p + \pi^+\nonumber
\end{eqnarray}
\begin{eqnarray}\label{channels_pi}
\nu_l(k)+n(p)&\rightarrow& l^{-}(k^{\prime})+\Delta^{+}(P)\\
           &&~~~~~~~~~~~~~~                     \searrow p + \pi^0\nonumber\\
           &&~~~~~~~~~~~~~~                       \searrow n  + \pi^+\nonumber
\end{eqnarray}
\begin{eqnarray}\label{channels_numubar_pi0}
\bar \nu_l(k)+p(p)&\rightarrow& l^{+}(k^{\prime})+\Delta^{0}(P)\\
           &&~~~~~~~~~~~~~~                     \searrow p + \pi^-\nonumber\\
           &&~~~~~~~~~~~~~~                       \searrow n  + \pi^0\nonumber
\end{eqnarray}
\begin{eqnarray}\label{channels_numubar_pi-}
\bar \nu_l(k)+n(p)&\rightarrow& l^{+}(k^{\prime})+\Delta^{-}(P)\\
           &&~~~~~~~~~~~~~~                     \searrow n + \pi^-\nonumber
\end{eqnarray}
Now we shall present the formalism for the incoherent and coherent pion productions in brief.
\subsection{INCOHERENT PION PRODUCTION}
In the case of incoherent pion production in $\Delta$ dominance model, the weak hadronic currents interacting with the nucleons in the nuclear medium excite a $\Delta$ resonance which decays into pions and nucleons. The pions interact with the nucleus inside the nuclear medium before coming out. The final state interaction of pions due to elastic, charge exchange scattering and the absorption of pions leads to reduction of pion yield. The nuclear medium effects on $\Delta$ properties lead to modifications in its mass and width which have been taken from the work of Oset et al.~\cite{Oset}.

In the case of incoherent one pion production process, the hadronic current $J^\mu$ for the $\Delta$ excitation from the proton target is given by
\begin{equation}\label{had_incoh}
J^\mu=\sqrt3 \bar\Psi_\alpha(p_{\Delta}) \mathcal O^{\lambda \mu} u(p)
\end{equation}
where ${\psi_\alpha}(p_{\Delta})$ and u(p) are the
Rarita Schwinger and Dirac spinors for the $\Delta$ and the nucleon, of
momenta $p_{\Delta}$ and p, respectively. $\mathcal O^{\lambda \mu}$ is the N-$\Delta$ transition operator given by $\mathcal O^{\lambda \mu}= \mathcal O^{\lambda \mu}_V + \mathcal O^{\lambda \mu}_A$ where
\begin{eqnarray}\label{mat_vector}
\mathcal O^{\lambda \mu}_V&=&\left(\frac{C^V_{3}(q^2)}{M}(g^{\alpha\mu}{\not q}-q^\alpha{\gamma^\mu}) + \frac{C^V_{4}(q^2)}{M^2}(g^{\alpha\mu}q \cdot p_{\Delta}-q^\alpha{p^{\mu}_\Delta})+\frac{C^V_5(q^2)}{M^2}(g^{\alpha\mu}q\cdot p-q^\alpha{p^\mu})+C^V_6(q^2) g^{\alpha\mu}\right)\gamma_5
\end{eqnarray}
and
\begin{eqnarray}\label{mat_axial}
\mathcal O^{\lambda \mu}_A&=&\frac{C^A_{3}(q^2)}{M}(g^{\alpha\mu}{\not q}-q^\alpha{\gamma^\mu}) + \frac{C^A_{4}(q^2)}{M^2}(g^{\alpha\mu}q \cdot p_{\Delta} - q^\alpha p^{\mu}_{\Delta}) + C^A_{5}(q^2)g^{\alpha\mu}+\frac{C^A_6(q^2)}{M^2}q^\mu q^\alpha 
\end{eqnarray}
where $p_{\Delta} = p+q$, $C^V_i$(i=3-6) are the vector and $C^A_i$(i=3-6) are the axial vector transition form factors.

The conserved vector current (CVC) hypothesis implies $C_6^V(q^2)$=0. The other form factors $C^V_i(i=3-5)$ are related in terms of the isovector electromagnetic form factors of the $p\rightarrow\Delta^+$ electromagnetic transition, and are determined from the analysis of data on photoproduction and electroproduction of $\Delta$. 
 
The N-$\Delta$ vector transition form factors given by Schreiner and von Hippel\cite{schvon} are 
\begin{eqnarray}\label{civ_schvon}
C_3^V(Q^2)&=&2.05~\left(1+\frac{Q^2}{M_V^2}\right)^{-2}\nonumber\\
C_4^V(Q^2)&=&-\frac{M}{M_\Delta}~C_3^V(Q^2)\nonumber\\
C_5^V(Q^2)&=&0
\end{eqnarray}
with $M_\Delta$ as the invariant mass of the $\pi$N system. 

The N-$\Delta$ axial vector transition form factors are given by\cite{schvon}
\begin{eqnarray}\label{dipole_ff}
C_i^A(Q^2)=C_i^A(0)~\left(1+\frac{Q^2}{M_A^2}\right)^{-2}\left(1-\frac{a_{i}Q^2}{(b_i+Q^2)}\right),~~~i=3,4,5
\end{eqnarray}
with $C_{3}^A(0)$~=~0, $C_{4}^A(0)$~=~-0.3, $C_{5}^A(0)$~=~1.2, $a_3=b_3$~=~0, $a_4=a_5$ = -1.21, $b_4=b_5~$=2 GeV$^2$, M$_A$~= 1.05 GeV.

While the various other parametrization for the  N-$\Delta$ transition form factor has been discussed in literature ~\cite{Leitner},\cite{Pasff},\cite{Lalakulich}. 

The parametrization given by Lalakulich et al.~\cite{Lalakulich} for the N-$\Delta$ transition form factors are given by
\begin{equation}\label{civ_lala}
C_i^V(Q^2)=C_i^V(0)~\left(1+\frac{Q^2}{M_V^2}\right)^{-2}~{\cal{D}}_i~~,~~~i=3,4,5.
\end{equation}
where
\begin{eqnarray}\label{di} 
{\cal{D}}_i&=&\left(1+\frac{Q^2}{4M_V^2}\right)^{-1}~~~\mbox{for}~~~i=3,4~~~ and\nonumber\\
{\cal{D}}_i&=&\left(1+\frac{Q^2}{0.776M_V^2}\right)^{-1}~~~~~~\mbox{for}~~~i=5.
\end{eqnarray} 
and 
\begin{eqnarray}\label{cia_lala}
C_i^A(Q^2)&=&C_i^A(0)~~\left(1+\frac{Q^2}{M_A^2}\right)^{-2}\left(1+\frac{Q^2}{3M_A^2}\right)^{-1},~~~i=3,4,5
\end{eqnarray}

Leitner et al.~\cite{Leitner} and Paschos et al.~\cite{Pasff} use the following form of the N-$\Delta$ vector transition form factor 
\begin{equation}\label{civ_pas_leit}
C_i^V(Q^2)=C_i^V(0)~\left(1+\frac{Q^2}{M_V^2}\right)^{-2}\left(1+\frac{Q^2}{4M_V^2}\right)^{-1}
\end{equation}
where $C_3^V(0)=1.95$, $C_4^V(0)=-\frac{M}{W}~C_3^V(0)$, $C_5^V(Q^2)=0$ with W as the center of mass energy ($\sqrt{(p+q)^2}$) and $M_\Delta$ as the mass of $\Delta$. For the axial vector part they use the same parametrization as used by Lalakulich et al.\cite{Lalakulich}. In the reactions given by Eq.\ref{chan_numu_pi+}-\ref{channels_numubar_pi-}, a $\Delta$ is produced, which subsequently decays into a nucleon and a pion, for example interaction of a neutrino with a proton inside the nucleus is given by $\nu_l(k) + p(p) \rightarrow l^-(k^\prime) + p(p^\prime) + \pi^+(k_\pi)$, for such a process the transition matrix element $\mathcal M_{fi}$ is given by
\begin{equation}\label{matrix_element}
\mathcal M_{fi}=\sqrt{3}\frac{G_F cos\theta_{c}}{\sqrt{2}}\frac{f_{\pi N \Delta}}{m_{\pi}} \bar u({\bf p}^{\prime}) k^{\sigma}_{\pi} {\mathcal P}_{\sigma \lambda} \mathcal O^{\lambda \mu} l_{\mu} u({\bf p})
\end{equation}
where $l_\mu$ is the leptonic current given by Eq.(\ref{lep_curr}), $\mathcal O^{\lambda \mu}$ is the N-$\Delta$ transition operator given by Eq.(\ref{mat_vector}) \& Eq.(\ref{mat_axial}), and ${\mathcal P}^{\sigma \lambda}$ is the $\Delta$ propagator in momentum space  which is given by~: 
\begin{equation}\label{prop}
{\mathcal P}^{\sigma \lambda}=\frac{{\it P}^{\sigma \lambda}}{P^2-M_\Delta^2+iM_\Delta\Gamma}
\end{equation}
with ${\it P}^{\sigma \lambda}$ as the spin-3/2 projection operator given as
\begin{eqnarray}\label{propagator}
{\it P}^{\sigma \lambda} = \sum_{spins} \psi^{\sigma} \bar \psi^{\lambda} = (\not P+M_{\Delta})
\left(g^{\sigma \lambda}-\frac{2}{3} \frac{P^{\sigma}P^{\lambda}}{M_{\Delta}^2}+\frac{1}{3}\frac{P^{\sigma} \gamma^{\lambda}-P^{\sigma} \gamma^{\lambda}}{M_{\Delta}}-\frac{1}{3}\gamma^{\sigma}\gamma^{\lambda}\right)
\end{eqnarray}

and the delta decay width $\Gamma$ is taken from~\cite{Oset} i.e.:
\begin{equation}\label{width}
\Gamma(W)=\frac{1}{6 \pi}\left(\frac{f_{\pi N \Delta}}{m_{\pi}}\right)^2 \frac{M}{W}|{\bf q}_{cm}|^3
\end{equation}
$|{\bf q}_{cm}|$ is the pion momentum in the rest frame of the resonance given by
\[|{\bf q}_{cm}|=\frac{\sqrt{(W^2-m_\pi^2-M^2)^2-4m_\pi^2M^2}}{2W}\]
and $M$ is the mass of nucleon and W is the center of mass energy.
In the nuclear medium the properties of $\Delta$ like its
mass and decay width $\Gamma$ to be used in Eq.(\ref{prop} ) are modified due
to the nuclear effects. These are mainly due to the following processes.

(i) In the nuclear medium $\Delta$s decay mainly through the $\Delta \rightarrow N\pi$ channel. The final nucleons have to be above the Fermi momentum $k_F$ of the nucleon in the nucleus thus inhibiting the decay as compared to the free decay of the $\Delta$ described by $\Gamma$ in Eq.(\ref{width}). This leads to a modification in the decay width of delta.  We have taken these modifications given by~\cite{Oset} where the modified delta decay width $\tilde\Gamma$ is given as 
\begin{equation}\label{modified_width}
\tilde\Gamma=\Gamma \times F(k_{F},E_{\Delta},k_{\Delta})
\end{equation}
 Here $F(k_{F},E_{\Delta},k_{\Delta})$ is the Pauli correction factor given by~\cite{Oset}:
\begin{equation}\label{pauli_correction}
F(k_{F},E_{\Delta},k_{\Delta})= \frac{k_{\Delta}|{{\bf q}_{cm}}|+E_{\Delta}{E^\prime_p}_{cm}-E_{F}{W}}{2k_{\Delta}|{\bf q^\prime}_{cm}|} 
\end{equation}
 $E_F=\sqrt{M^2+k_F^2}$, $k_{\Delta}$ is the $\Delta$ momentum and  $E_\Delta=\sqrt{W+k_\Delta^2}$. 

(ii) In the nuclear medium there are additional decay channels open
due to two and three body absorption processes like $\Delta N
\rightarrow N N$ and $\Delta N N\rightarrow N N N$ through which
$\Delta$ disappears in the nuclear medium without producing a pion,
while a two body $\Delta$ absorption process like $\Delta N
\rightarrow \pi N N$ gives rise to some more pions. These nuclear
medium effects on the $\Delta$ propagation are included by describing
the mass and the decay width in terms of the self energy of
$\Delta$. These considerations lead to the following modifications in the width $\tilde\Gamma$ and mass $M_\Delta$ of the $\Delta$ resonance. 
\begin{equation}\label{final_width}
\frac{\tilde\Gamma}{2}\rightarrow\frac{\tilde\Gamma}{2} - Im\Sigma_\Delta~~\text{and}~~
M_\Delta\rightarrow M_\Delta + Re\Sigma_\Delta.
\end{equation}
 The expressions for the real and the imaginary parts of $\Sigma_\Delta$ are~\cite{Oset}:
\begin{eqnarray}\label{real_imag_selfenergy}
Re{\Sigma}_{\Delta}&=&40 \frac{\rho}{\rho_{0}}MeV ~~and \nonumber\\
-Im{{\Sigma}_{\Delta}}&=&C_{Q}\left (\frac{\rho}{{\rho}_{0}}\right )^{\alpha}+C_{A2}\left (\frac{\rho}{{\rho}_{0}}\right )^{\beta}+C_{A3}\left (\frac{\rho}{{\rho}_{0}}\right )^{\gamma}~~~~
\end{eqnarray}
In the above equation $C_{Q}$ accounts for the $\Delta N  \rightarrow
\pi N N$ process, $C_{A2}$ for the two-body absorption process $\Delta
N \rightarrow N N$ and $C_{A3}$ for the three-body absorption process $\Delta N N\rightarrow N N N$. The coefficients $C_{Q}$, $C_{A2}$, $C_{A3}$ and $\alpha$, $\beta$ and $\gamma$ are taken from Ref.~\cite{Oset}.

Thus, in the local density approximation the expression for the total cross section for the neutrino induced charged current 1$\pi^+$ production from proton target is written as
\begin{eqnarray}\label{sigma_inelas}
\sigma_A(E)&=& \frac{1}{(4\pi)^5}\int_{r_{min}}^{r_{max}}\rho_{p}(r) d\vec r\int_{Q^{2}_{min}}^{Q^{2}_{max}}dQ^{2}
\times \int^{k^\prime_{max}}_{k^\prime_{min}} d{k^\prime} \int_{-1}^{+1}dcos\theta_{\pi } \nonumber\\
&&\times \int_{0}^{2\pi}d\phi_{\pi} 
\frac{\pi|\vec  k^\prime||\vec k_{\pi}|}{M E_{\nu}^2 E_{l}}\frac{1}{E_{p}^{\prime}+E_{\pi}\left(1-\frac{|\vec q|}{|\vec k_{\pi}|}cos\theta_{\pi }\right)}\bar\Sigma \Sigma|\mathcal M_{fi}|^2
\end{eqnarray}
Similar expression is obtained for the cross section for neutrino induced charged current 1$\pi^+$ production from neutron target with $\rho_{p}$ replaced by $\frac{1}{9}\rho_{n}$. A factor of $\frac{1}{9}$ comes with $\rho_n$ due to the Clebsch Gordan coefficient occurring in the production of $\pi^+$ from the neutron target ($\nu_\mu + n \rightarrow \mu^- + \Delta^+, \Delta^+ \rightarrow n + \pi^+) $ as compared to the $\pi^+$ production from the proton target. For the antineutrino induced $1\pi^-$ production process the factor of $\frac{1}{9}$ will come with $\rho_{p}$ i.e. $\pi^-$ would get produced dominantly from neutron target.

The pions produced in these processes inside the nucleus may
rescatter, produce more pions or may get absorbed while coming
out from the final nucleus. These are treated using Monte Carlo simulations by generating a pion of given momentum and charge at a point {\bf r} in the nucleus. Assuming the real part of the pion nuclear potential to be weak as compared with their kinetic energies, they are propagated following straight lines till they are out of the nucleus. At the beginning, the pions are placed at a point (${\bf r}={\bf b},z_{in}$), where $z_{in}= -\sqrt{{\bf R^2}-|{\bf b}|^2}$, with {\bf b} as the random impact parameter, obeying $|{\bf b}|<R$. R is upper bound for the nuclear radius, which is chosen to be such that $\rho(R)\approx 10^{-3} \rho_{0}$, with $\rho_{0}$ is the normal nuclear matter density. The pion is then made to move along z-direction in small steps until it comes out of the nucleus. We have taken the results of Vicente Vacas~\cite{Private} for the final state interaction of pions which has been discussed in Ref.~\cite{Vicente}.

\subsection{COHERENT PION PRODUCTION}
The $\nu_{\mu}$ induced coherent one pion production on $^{16}O$ target is given by $\nu_{\mu} + _{8}^{16}O \rightarrow \mu^{-} + _{8}^{16}O + \pi^{+}$ for which the cross section is given by Eq.(\ref{sigma_inelas}). 
However, the matrix element $\mathcal M_{fi}$ is now given by 
\begin{equation}\label{matrix_coh}
\mathcal M_{fi} =\frac{G_{F}}{\sqrt{2}} cos\theta_{c} l^{\mu} J_{\mu} {\cal F}(\vec q - \vec k_{\pi})
\end{equation}

where $l^{\mu}$ is the leptonic current given by Eq.(\ref{lep_curr}) and $J_{\mu}$ is the hadronic current given by~\cite{Coh1} 
\begin{equation}\label{coh_had_curr}
J_{\mu}= \sqrt{3} \frac{f_{\pi N \Delta}}{m_{\pi}} \sum _{r,s} {\bar u_{s}}(p) k_{\pi \sigma}\mathcal P^{\sigma \lambda} \mathcal O_{\lambda \mu} u_{r}(p)
\end{equation}
$P^{\sigma \lambda}$ is given by Eq.(\ref{prop}),  $\mathcal O_{\lambda \mu}$ is given by Eqs. (\ref{mat_vector}) and (\ref{mat_axial}), and u(p) is Dirac spinor for the nucleons.

Here ${\cal F}(\vec q - \vec k_{\pi})$ is the nuclear form factor, given by 
\begin{eqnarray}\label{ff}
{\cal F}(\vec q-\vec k_\pi)=\int d^{3}{\vec r} \left[{\rho_p ({\vec r})}+\frac{1}{3}{\rho_n ({\vec r})}\right]e^{-i({\vec q}-{\vec k}_\pi).{\vec r}}
\end{eqnarray}
When pion absorption effects are taken into account using the Eikonal approximation then the nuclear form factor ${\cal F}({\vec q}-{\vec k_\pi})$ is modified to $\tilde{\cal F}({\vec q}-{\vec k_\pi})$, which is calculated in Eikonal approximation to be~\cite{Carassco}:
\begin{equation}\label{mod_ff}
\tilde{\cal F}({\vec q}-{\vec k_\pi})=2\pi\int_0^\infty b~db\int_{-\infty}^\infty dz~\rho({\vec b}, z)~J_0(k_\pi^tb)~~e^{i(|{\vec q}|-k_\pi^l)z} e^{-if({\vec b}, z)}
\end{equation} 
where 
\[f({\vec b}, z)=\int_z^{\infty} \frac{1}{2|{\vec{k}_\pi}|}{\Pi(\rho({\vec b}, z^\prime))}dz^\prime\]
$k_{\pi}^{l}$ and $k_{\pi}^{t}$ are the longitudinal and transverse components of the pion momentum and $\Pi$ is the self-energy of pion, the expression for which is taken from Ref.~\cite{Carassco} and is given by 
\begin{equation}\label{self_energy}
\Pi(\rho({\vec b}, z^\prime))=\frac{4}{9}\left(\frac{f_{\pi N\Delta}}{m_\pi}\right)^2\frac{M^2}{W^2}|{\vec k_\pi}|^2~\rho({\vec b}, z^\prime)~\frac{1}{W-{\tilde M}_\Delta+\frac{i{\tilde \Gamma}}{2}}
\end{equation}
 
Using the matrix element given by Eq.(\ref{matrix_coh}) and the modifications in the $\Delta$ mass and width in the nuclear medium given by Eq.(\ref{final_width}) we calculate the total scattering cross section $\sigma$ given in Eq.(\ref{sigma_inelas}).

\begin{figure*}
\includegraphics{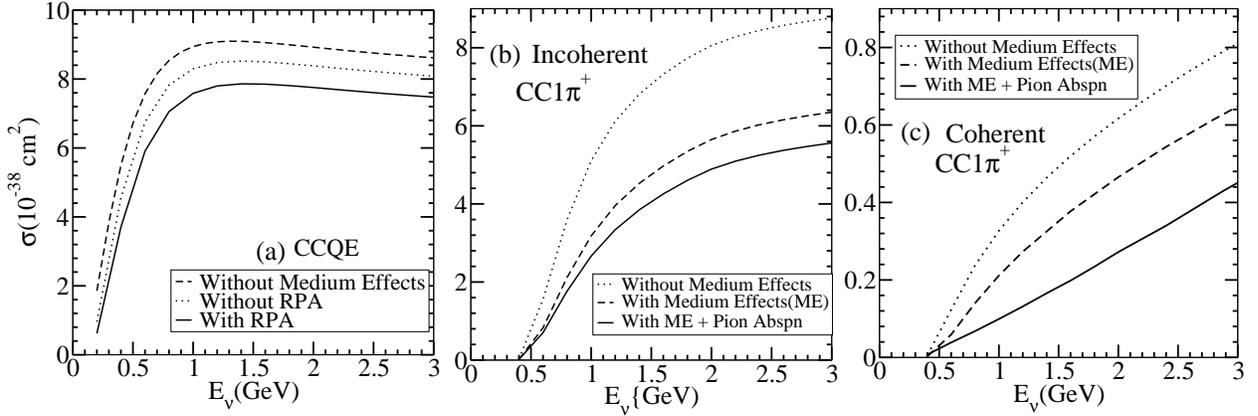}
\caption{Total scattering cross section($\sigma$) as a function of neutrino energy for the charged current (a) Quasielastic (b) Incoherent and (c) Coherent processes for $\nu_{\mu}$ induced reaction in $^{16}O$.}
\label{fg:fig3}
\end{figure*}
\begin{figure*}
\includegraphics{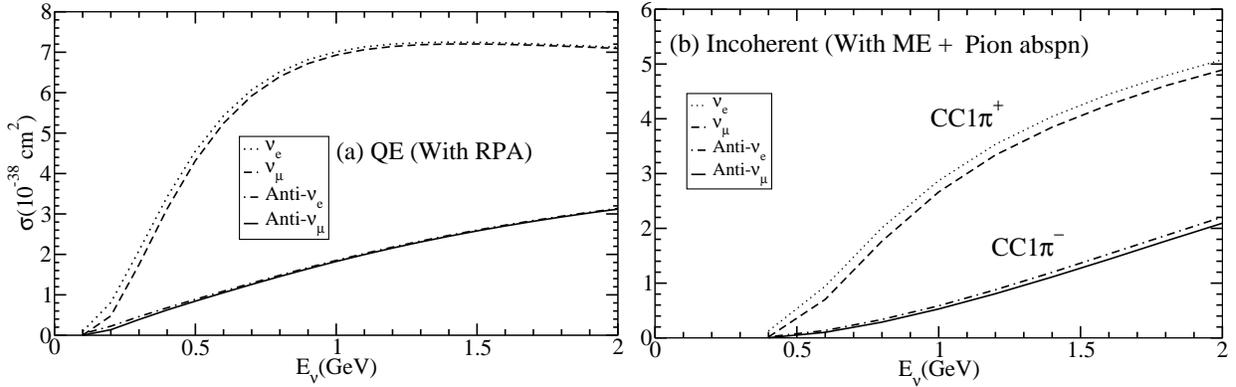}
\caption{Total scattering cross section($\sigma$) as a function of neutrino energy for the charged current (a) Quasielastic and (b) Incoherent processes for $\nu_e$, $\nu_{\mu}$, $\bar \nu_e$ and $\bar \nu_{\mu}$ induced reaction in $^{16}O$. These results are presented for $\sigma$ calculated with RPA in CCQE scattering (Fig.4(a)) and with medium and pion absorption effects in the case of CC1$\pi$ production processes (Fig.4(b)).}
\label{fg:fig4}
\end{figure*}
\begin{figure*}
\includegraphics{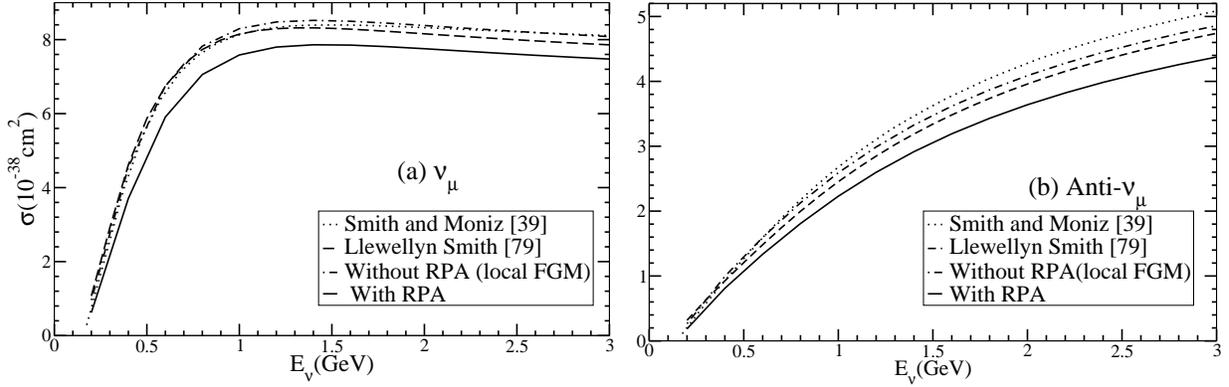}
\caption{Total scattering cross section($\sigma$) as a function of neutrino energy for $\nu_{\mu}$ ($\bar \nu_{\mu}$) induced reaction in $^{16}O$. These results have been presented for the cross sections calculated by using different models.}
\label{fg:fig5}
\end{figure*}

\begin{figure*}
\includegraphics{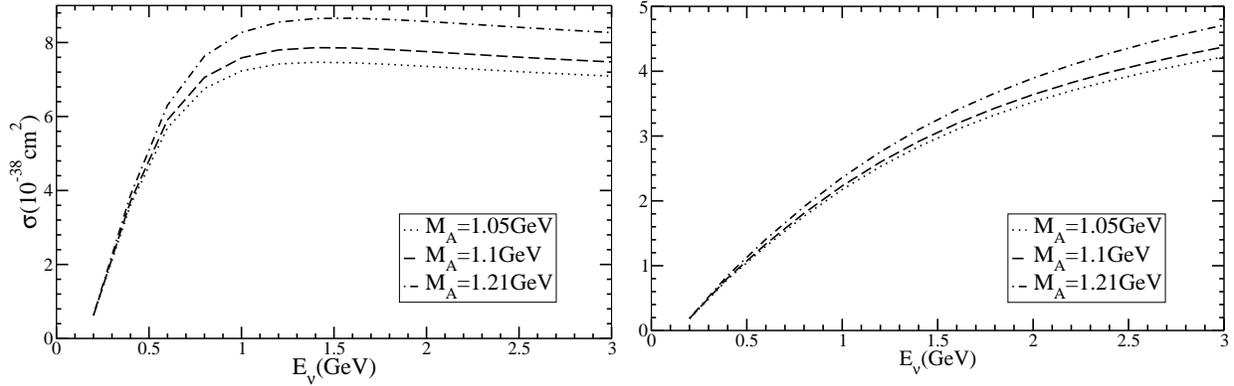}
\caption{Total scattering cross section($\sigma$) as a function of neutrino energy for the charged current quasielastic lepton production process induced by (a) $\nu_{\mu}$ and (b) $\bar \nu_{\mu}$ in $^{16}O$ in the local FGM with RPA effects. The present curves show the $M_{A}$ dependence on the cross section.}
\label{fg:fig6}
\end{figure*}
\begin{figure*}
\includegraphics{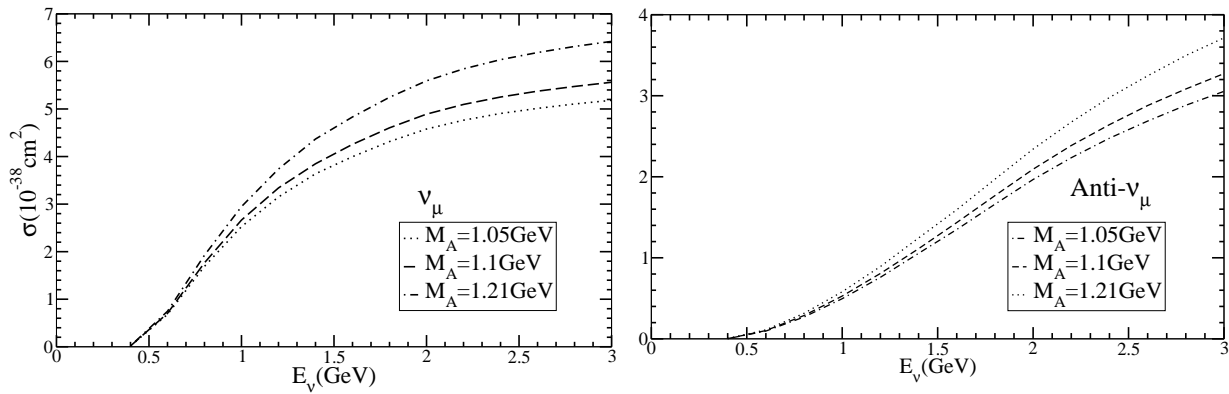}
\caption{Total scattering cross section($\sigma$) as a function of neutrino(antineutrino) energy for the charged current incoherent $1\pi^+$($1\pi^-$) production from $^{16}O$ target with nuclear medium and final state interaction effects. The present curves show the $M_{A}$ dependence on the cross section.}
\label{fg:fig7}
\end{figure*}
\section{Results and Discussion}
\subsection{Total Scattering Cross Section $\sigma$}
For the charged current quasielastic(CCQE) reaction, the numerical results are obtained from Eq.(\ref {cross_section_quasi}) using the expression for the form factors given by Bradford et al.~\cite{BBBA05} with  vector dipole mass ${M}_{V}$=0.84GeV and axial dipole mass $M_A$=1.1GeV. In the case of charged current induced incoherent and coherent pion productions the results for the total cross sections are obtained from Eq.(\ref{sigma_inelas}) using the matrix elements given in Eq.(\ref{matrix_element}) \& Eq.(\ref{matrix_coh}) respectively and the N-$\Delta$ transition form factors given by Lalakulich et al.~\cite{Lalakulich} given by Eq.(\ref{civ_lala}) and Eq.(\ref{cia_lala}) with $M_{A}$=1.1 GeV. In Figs.\ref{fg:fig3}a-c, we have shown the results with nuclear medium effects for the total cross section $\sigma$ for CCQE and incoherent \& coherent CC1$\pi^+$ production cross sections in $^{16}$O, for $\nu_\mu$ induced reaction. 

For the quasielastic process, in the case of charged current $\nu_\mu$ induced lepton production cross section the results have been presented for the cross section calculated for the free case, with nuclear medium effects without RPA i.e. our local Fermi gas model, and with nuclear medium effects including RPA. These results have been shown in Fig.(\ref{fg:fig3}a). We find that when the cross section is calculated in the local Fermi gas model the reduction in the cross section is around $18\%$ at $\text E_{\nu_\mu}$=0.4GeV and around $10\%$ at $\text E_{\nu_\mu}$=1-3GeV from the cross sections calculated for the free case. However, when we encorporate the RPA effects, there is further reduction in the cross section which is about $30\%$ at $\text E_{\nu_\mu}$=0.4GeV, $15\%$ at $E_{\nu_\mu}$=1.0GeV and around $12\%$ at $E_{\nu_\mu}$=2-3GeV. 

The numerical results for the $\nu_\mu$ induced incoherent 1$\pi^+$ production process have been shown in Fig.(\ref{fg:fig3}b) and we find that the nuclear medium effects lead to a reduction of around 40$\%$ for $\text E_{\nu_\mu}$=0.4~GeV, 30$\%$ for $\text E_{\nu_\mu}$=1.5~GeV and 28$\%$ for $\text E_{\nu_\mu}$=3~GeV. When pion absorption effects are also taken into account along with the nuclear medium effects there is a further reduction in the cross section which is around 15$\%$ for $\text E_{\nu_\mu}$=1.0~GeV and 12$\%$ for $\text E_{\nu_\mu}$=3~GeV. For $\nu_\mu$ induced coherent 1$\pi^+$ production process the numerical results have been shown in Fig.(\ref{fg:fig3}c) and we find that the nuclear medium effects lead to a reduction of around 40$\%$ for $\text E_{\nu_\mu}$=0.8~GeV, 24$\%$ for $\text E_{\nu_\mu}$=2.0~GeV and 15$\%$ at $\text E_{\nu_\mu}$=3~GeV. When pion absorption effects are also taken into account along with the nuclear medium effects there is a further reduction in the cross section which is around 50$\%$ for $\text E_{\nu_\mu}$=0.8~GeV, 35$\%$ for $\text E_{\nu_\mu}$=2.0~GeV and 25$\%$ $\text E_{\nu_\mu}$=3~GeV.

Thus, in the case of incoherent production of pions, the reduction due to nuclear medium effects in the production process is larger than the reduction due to final state interaction while in the case of coherent pion production, the reduction due to final state interaction is quite large as compared to the reduction due to the nuclear medium effects.
Furthermore, the contribution of the cross section calculated in the case of coherent pion production process in the nuclear medium with final state interaction effects is around 6-7$\%$ to the total (incoherent+coherent) one pion production cross section in the energy region of $0.4GeV<E<3GeV$, and due to this we have not discussed the form factor dependence, $Q^2$ distribution, etc. in the results presented here for the coherent pion production process as their contribution to the total lepton events in the sub-GeV energy region of present interest is not very significant. Our results for the coherent process also agrees with the other recent calculations performed by the  various groups~\cite{Sehgal},\cite{Nieves},\cite{Ruso}.
 
In Figs.(\ref{fg:fig4}a) and (\ref{fg:fig4}b), we have presented the results for the total scattering cross section $\sigma$ as a function of neutrino energy $E_\nu$ for the charged current lepton production process induced by neutrino(antineutrino) in $^{16}O$ in the case of quasielastic and incoherent 1$\pi$ production processes. These results have been presented for $\nu_e$, $\nu_\mu$, ${\bar\nu}_e$ and ${\bar\nu}_\mu$ in the local Fermi gas model with RPA effects in the case of quasielastic process and with nuclear medium and final state interaction effects in the case of incoherent 1$\pi^+$(neutrino) and 1$\pi^-$(antineutrino) production  processes. 

In Figs.(\ref{fg:fig5}a) and (\ref{fg:fig5}b), we have compared our results for the $\nu_\mu$ and ${\bar\nu}_\mu$ induced charged current quasielastic lepton production cross sections in $^{16}O$, obtained in the local Fermi gas model with and without the RPA effects, with the results obtained in the Fermi gas model given by Smith and Moniz~\cite{SmithMoniz} and Llewellyn Smith ~\cite{lsmith}, which have been used in some of the Monte Carlo generators. We find that our results in the local Fermi gas model are in fairly good agreement (within 2$\%$) with their results~\cite{SmithMoniz,lsmith} in the case of $\nu_\mu$ induced process, while in the case of ${\bar\nu}_\mu$ process the results obtained in our local Fermi gas model are within 3-4$\%$ with the results obtained by using Llewellyn Smith's~\cite{lsmith} Fermi gas model, however, the results obtained by Smith and Moniz~\cite{SmithMoniz} Fermi gas model is about 5-6$\%$ higher. We have also compared our results in the local FGM with the non-relativistic FGM of Gaisser and O' Connell~\cite{Gaisser}(not shown here) and found the cross sections to be within 2-3 $\%$ at the neutrino energies of the present interest. When RPA correlation effects are taken into account the cross section decreases. We find that our results for the total scattering cross section $\sigma(E)$ in the case of charged current neutrino induced process calculated in the local FGM with RPA effects agree with the recent calculations performed by Leitner et al.~\cite{Leitner1}, Benhar et al.\cite{Benhar} and Nieves et al.~\cite{Nieves1}.

We have shown in Figs.(\ref{fg:fig6}a) and (\ref{fg:fig6}b), the effect of varying the axial dipole mass $M_A$ on the total scattering cross section $\sigma$ for the $\nu_\mu$ and ${\bar\nu}_\mu$ induced charged current quasielastic reaction cross sections in $^{16}O$, obtained in the local Fermi gas model with RPA effects using Bradford et al.~\cite{BBBA05}  parametrization of the weak form factors given in Eq.(\ref{ge_gm_bbba01}) with  vector dipole mass ${M}_{V}$=0.84GeV. We find that $\sigma$ increases by about 5$\%$ at $E_{\nu_\mu}$=0.5GeV and by 10$\%$ at $E_{\nu_\mu}$=2-3GeV when $M_A$ is taken as 1.21GeV, while it decreases by about 3$\%$ at $E_{\nu_\mu}$=0.5GeV and by 5$\%$ at $E_{\nu_\mu}$=2-3GeV when $M_A$=1.05GeV as compared to the cross sections calculated with $M_A$=1.1GeV. While in the case of ${\bar\nu}_\mu$ induced reaction cross section the dependence on $M_A$ becomes small which is around 6$\%$ at $E_{\bar \nu_\mu}$=1-3GeV when $M_A$ is taken as 1.21GeV, while the decrease is about 3$\%$ at $E_{\bar \nu_\mu}$=1-3GeV when $M_A$=1.05GeV as compared to the cross sections calculated with $M_A$=1.1GeV.

We have studied the Landau-Migdal parameter ($g^{\prime}$) dependence(not shown here) given in Eq.\ref{longi_part}, expression for which has been used in calculating the cross section with RPA effects, on the total cross section. We find that a 10$\%$ uncertainty in g$^\prime$ leads to a 5-6$\%$ of uncertainty in the cross section. Also we have studied the nuclear density dependence on the total scattering cross section(not shown here). Using other densities(modified harmonic oscillator or 3 parameter Fermi density) leads to 2-3$\%$ of  uncertainty in the cross section.

In Figs.(\ref{fg:fig7}a) and (\ref{fg:fig7}b), we have shown the effect of varying the axial dipole mass $M_A$ on the total scattering cross section $\sigma$ for the $\nu_\mu$ and ${\bar\nu}_\mu$ induced 1$\pi$ production cross sections in $^{16}O$ with nuclear medium and final state interaction effects. We have used the N-$\Delta$ transition form factor parameterizations given by Lalakulich et al.~\cite{Lalakulich}. We find that $\sigma$ increases by about 5$\%$ at $E_\nu$=0.5GeV and by 12$\%$ at $E_\nu$=2-3GeV when $M_A$ is taken as 1.21GeV, while it decreases by about 3$\%$ at $E_\nu$=0.5GeV and by 6$\%$ at $E_\nu$=2-3GeV when $M_A$=1.05GeV as compared to the cross sections calculated with $M_A$=1.1GeV. While in the case of ${\bar\nu}_\mu$ induced reaction cross section the dependence on $M_A$ becomes small which is around 10$\%$ at $E_{{\bar\nu}_\mu}$=1-3GeV when $M_A$ is taken as 1.21GeV, while the decrease is about 6$\%$ at $E_\nu$=1-3GeV when $M_A$=1.05GeV as compared to the cross sections calculated with $M_A$=1.1GeV. 

Our results for the total scattering cross section $\sigma(E)$ in the case of charged current neutrino induced incoherent pion production process in the $\Delta$ dominance model calculated with nuclear medium and final state interaction effects agree with the numerical results of Leitner et al.~\cite{Leitner1} and Benhar et al.\cite{Benhar}.

\subsection{Differential Scattering Cross Section $<\frac{d{\sigma}}{dQ^2}>$}

In this section we shall present the results for the flux averaged differential scattering cross section $<\frac{d{\sigma}}{dQ^2}>$ as a function of $Q^2$. This has been obtained by integrating $\frac{d{\sigma}}{dQ^2}$ over the atmospheric neutrino flux given by Honda et al.~\cite{Honda2} for the SuperK site. 

The flux averaged differential scattering cross section $<\frac{d{\sigma}}{dQ^2}>$ is defined as
\begin{equation}\label{avg_diff_xsec}
<\frac{d{\sigma}}{dQ^2}>=\frac{\int_{{E_\nu}_{min}}^{{E_\nu}_{max}} \frac{d{\sigma}}{dQ^2} \phi(E) dE}{\int_{{E_\nu}_{min}}^{{E_\nu}_{max}} \phi(E) dE}
\end{equation}
where $\frac{d{\sigma}}{dQ^2}$ is the differential scattering cross section for the $Q^2$ distribution and $\phi(E)$ is the atmospheric neutrino flux. 

We discuss the nuclear medium modification effects on $Q^2$-distribution, the effect of $M_A$ and different parameterizations of the various isovector form factors in the case of quasielastic process and N-$\Delta$ transition form factors in the case of 1$\pi$ production process on $Q^2$ distribution. In addition to these, we shall also present the results to show the dependence of $Q^2$ distribution for the two fluxes Kam1997 and Kam2000 given by Honda~\cite{Honda2}.

\begin{figure*}
\includegraphics{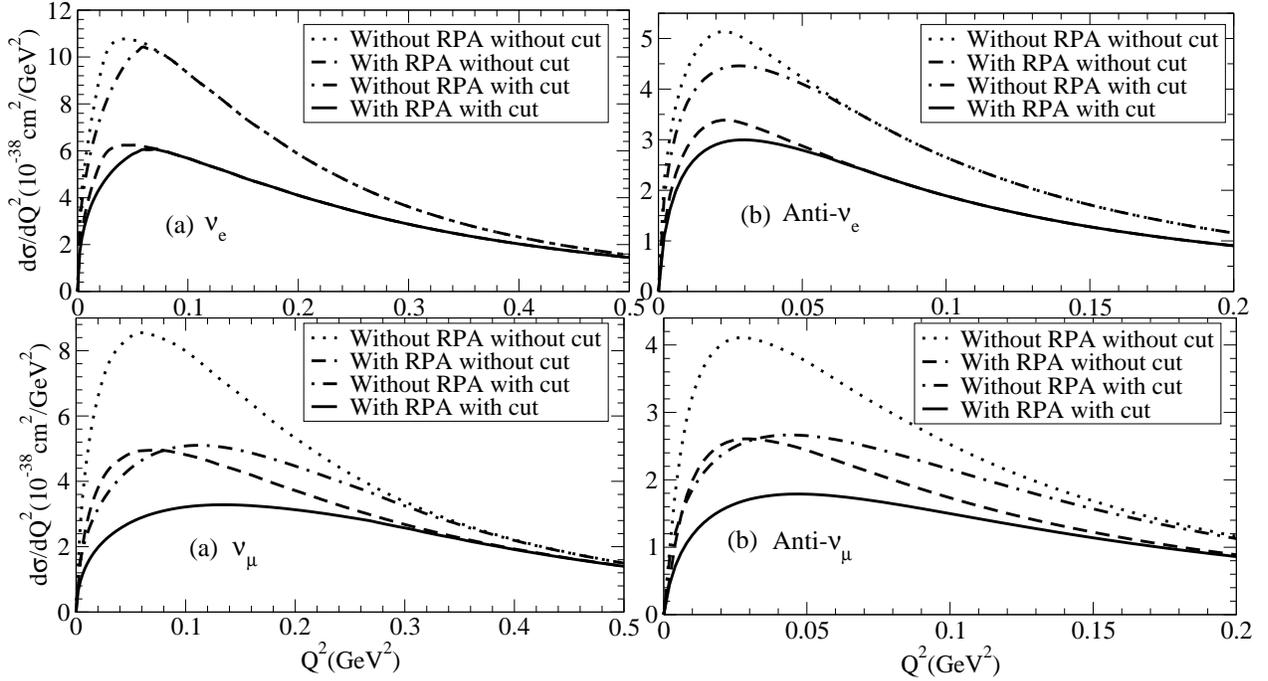}
\caption{$<\frac{d\sigma}{dQ^2}>$ vs $Q^2$ for the quasielastic process induced by electron type(upper panels) and muon type (lower panels) (a) neutrino and (b) antineutrino in $^{16}O$. The results are presented for the $Q^{2}$-distribution with and without applying cuts on the lepton's energy and momenta. The dashed(dotted) line is the result in the local FGM with(without) RPA effects and without cuts. The solid(dashed-dotted) line is the result in the local FGM with(without) RPA effects and with cuts ($E_{l} < 1.33GeV$, $p_{e} \geq 100MeV$ and $p_{\mu} \geq 200MeV$).}
\label{fg:fig8}
\end{figure*}

In Figs.(\ref{fg:fig8}a) and (\ref{fg:fig8}b), we present the results for the $Q^2$-distribution in the case of charged current quasielastic lepton production process induced by electron and muon neutrino(antineutrino). The results have been presented for the $Q^2$-distribution calculated in the local Fermi gas model with and without RPA effects. In the case of $\nu_e$ and $\nu_\mu$ induced processes the results have been shown in Fig.(\ref{fg:fig8}a). We find that in the case of $\nu_e$ induced process the differential cross section is calculated in the local Fermi gas model with RPA effects the reduction in the cross section is around 42$\%$ in the peak region of $Q^2$(=0.044GeV$^2$) and around 30$\%$ at $Q^2$=0.2GeV$^2$ as compared to the cross section calculated without the RPA effects. For ${\bar\nu}_e$, there is a shift in the peak region which is towards low $Q^2$(=0.022GeV$^2$) and the reduction is around 35$\%$ which is smaller than in the case of $\nu_e$ induced process and at high $Q^2$ (=0.1GeV$^2$) the reduction is around 30$\%$ as compared to the cross section calculated without the RPA effects. When a cut on the electron's energy ($E_e < 1.33GeV$) and momenta($p_e \geq 100MeV$) is applied, then there is a small change in the $Q^2$ spectrum which in turn leads to a small change in the event rates.

In the case of $\nu_\mu$ induced process the reduction in the cross section  when RPA effects are taken into account is around 40$\%$ in the peak region of $Q^2$(=0.06GeV$^2$) and around $30\%$ at $Q^2$=0.2GeV$^2$ as compared to the cross section calculated without the RPA effects in the local Fermi gas model.  In the case of ${\bar\nu}_\mu$, there is a shift in the peak region which is towards low $Q^2$(=0.028GeV$^2$) and the reduction is around $36\%$ which is smaller than in the case of $\nu_\mu$ induced process and at high $Q^2$ (=0.2GeV$^2$) the reduction is around $22\%$ as compared to the cross section calculated without the RPA effects. When a cut on the muon's energy ($E_{\mu} < 1.33GeV$) and momenta($p_{\mu} \geq 200MeV$) are applied, then there is a large suppression in the $Q^2$ distribution in the peak region as shown in the Fig.\ref{fg:fig8}. We find that the inclusion of RPA effects in our local Fermi gas model with a cut on muon's momenta and energy results in a large suppression in the event rates in comparison to the muon events calculated by using the local Fermi gas model without RPA effects and without applying cuts on the muon's energy and momenta.

\begin{figure*}
\includegraphics{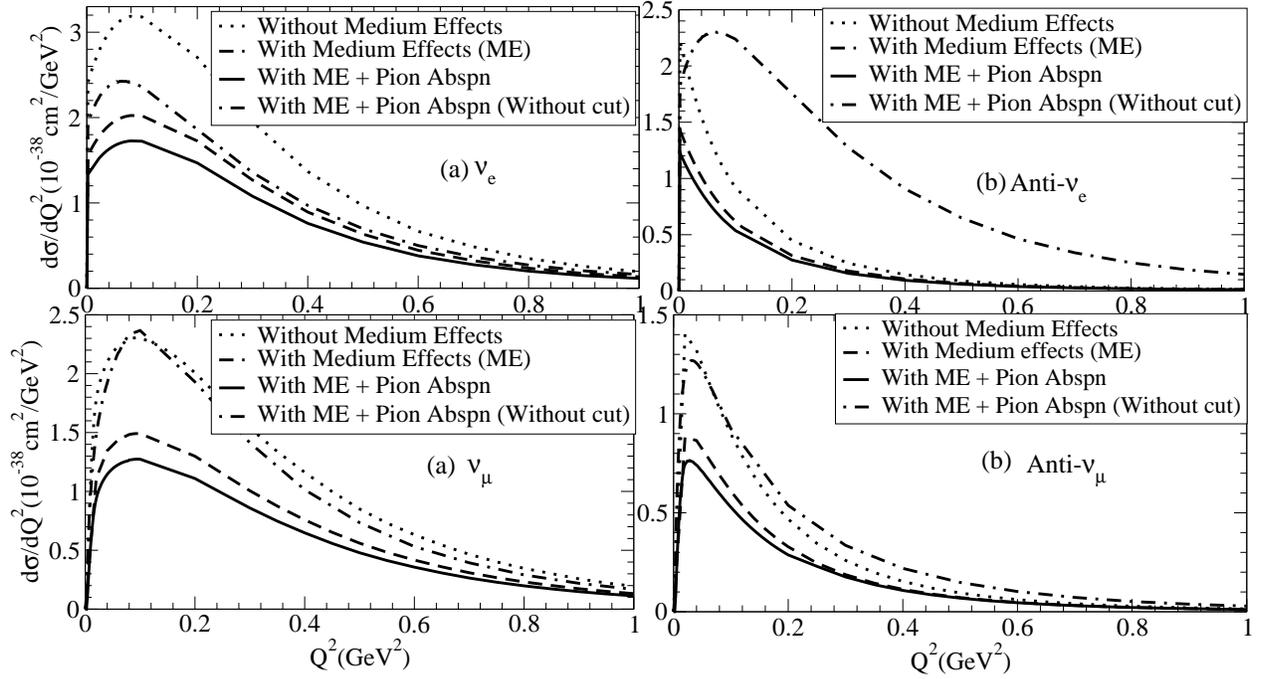}
\caption{$<\frac{d\sigma}{dQ^2}>$ vs $Q^2$ for the incoherent process induced by electron type(upper panels) and muon type (lower panels) (a) neutrino and (b) antineutrino. The results are presented for the $Q^2$ distribution with cuts on the lepton energy($E_{l} < 1.33GeV$) and momenta($p_{e} \geq 100MeV$ and $p_{\mu} \geq 200MeV$) calculated with(without) nuclear medium and nuclear medium \& final state interaction effects. The results for the $Q^2$ distribution with nuclear medium \& final state interaction effects but without putting cut on lepton's momenta and energy have also been presented here(dashed-dotted line).}
\label{fg:fig9}
\end{figure*}

The $Q^2$ distribution in the case of charged current electron and muon neutrino(antineutrino) induced incoherent 1$\pi^+$(1$\pi^-$) production processes have been shown in Figs.(\ref{fg:fig9}a) and (\ref{fg:fig9}b). The results have been shown without and with the effect of nuclear medium as well as with the final state interaction of pions taken into account along with the nuclear medium effects. In the case of $\nu_e$ and $\nu_\mu$ induced processes the results have been shown in Fig.(\ref{fg:fig9}a). We find that in the case of $\nu_e$ induced process when the cut is applied on the electron's energy($E_e < 1.33GeV$) and momenta($p_e \geq 100MeV$), and the nuclear medium effects are also taken into account, the reduction in the cross section is around 36$\%$ in the peak region of $Q^2$(=0.08GeV$^2$) and around 33$\%$ at $Q^2$=0.7GeV$^2$ as compared to the cross section calculated without taking nuclear medium effects into account. When final state interactions effects are also taken into account there is a further reduction of around 15$\%$ in the peak region and around 14$\%$ at $Q^2$=0.7GeV$^2$. For ${\bar\nu}_e$, there is a shift in the peak region which is towards low $Q^2$(=0.002GeV$^2$) and the reduction is around 35$\%$ and at high $Q^2$ (=0.2GeV$^2$) the reduction is around 30$\%$ as compared to the cross section calculated without medium effects. Here we have also shown the results of the $Q^2$ distribution calculated by including nuclear medium and final state interaction effects but without applying any cuts on lepton energy and momenta, it has been found that this results in an enhancement in the distribution particularly in the peak region of $Q^2$.

In the case of $\nu_{\mu}$ induced process, when the differential cross section is calculated by applying cuts on the lepton's energy and momenta, with the nuclear medium effects taken into account, the reduction in the cross section is around $35\%$ in the peak region of $Q^2$($\approx$0.1GeV$^2$) as compared to the cross section calculated without the nuclear medium effects. When pion absorption effect is also taken into account there is further reduction of about $15\%$. In the case of ${\bar\nu}_\mu$ induced process there is a shift in the peak region which is towards low $Q^2$(=0.02GeV$^2$) and the nature of reduction is almost the same as in the case of $\nu_\mu$ induced process. When there is no cut applied on the muon's energy and momenta, then there is a change in the nature of reduction. For example, in the case of $\nu_{\mu}$ induced process the reduction in the cross section calculated with the nuclear medium effects taken into account, is around 35$\%$ in the peak region of $Q^2$ as compared to the differential cross section calculated without the nuclear medium effects, and when pion absorption effect is also taken into account there is further reduction of about 14$\%$, while for $\bar \nu_{\mu}$ induced process this reduction is around 30$\%$ in the peak region and with medium and final state interaction  effects taken into account, the further reduction in the differential cross section is around 12$\%$. To see the effect of applying cuts on the muon's energy and momenta, we calculate $Q^2$ distribution with medium effects and final state interaction effects and find that when cuts are taken into account in the case of $\nu_{\mu}$ induced reaction the reduction is around 45$\%$ in the peak region and becomes 32$\%$ around $Q^2=$0.7$ GeV^2$. However, in the case of $\bar \nu_{\mu}$ induced reaction this reduction is around 35$\%$ in the peak region and becomes 24$\%$ around $Q^2=$0.2$ GeV^2$. 

\begin{figure*}
\includegraphics{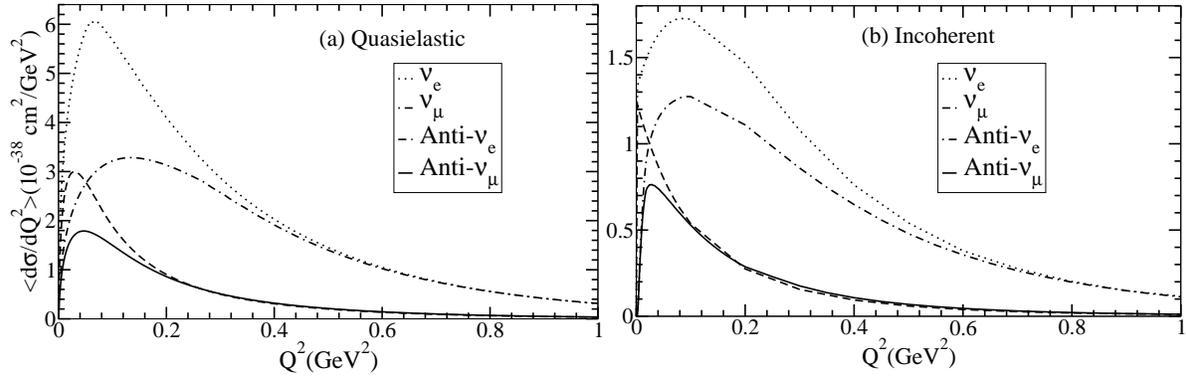}
\caption{$<\frac{d\sigma}{dQ^2}>$ vs $Q^2$ for (a)Quasielastic and (b)Incoherent processes induced by different flavors of (a) neutrino and (b) antineutrino averaged over the atmospheric neutrino flux given by Honda et al. \cite{Honda1},\cite{Honda2}. The results have been presented after putting cut on lepton momenta and energy.}
\label{fg:fig10}
\end{figure*}
In Fig.(\ref{fg:fig10}a), we show the results for the $Q^2$ distribution in the case of charged current quasielastic lepton production process induced by $\nu_e$, $\bar\nu_e$, $\nu_\mu$ and $\bar\nu_\mu$ in $^{16}$O calculated in the local Fermi gas model with RPA effects.  The results for the $Q^2$ distribution in the case of charged current incoherent 1$\pi^+$ production process induced by $\nu_e$, $\nu_\mu$, and 1$\pi^-$ production process induced by $\bar\nu_e$, $\bar\nu_\mu$ in $^{16}$O with nuclear medium and pion absorption effects have been shown in Fig.(\ref{fg:fig10}b).
\begin{figure*}
\includegraphics{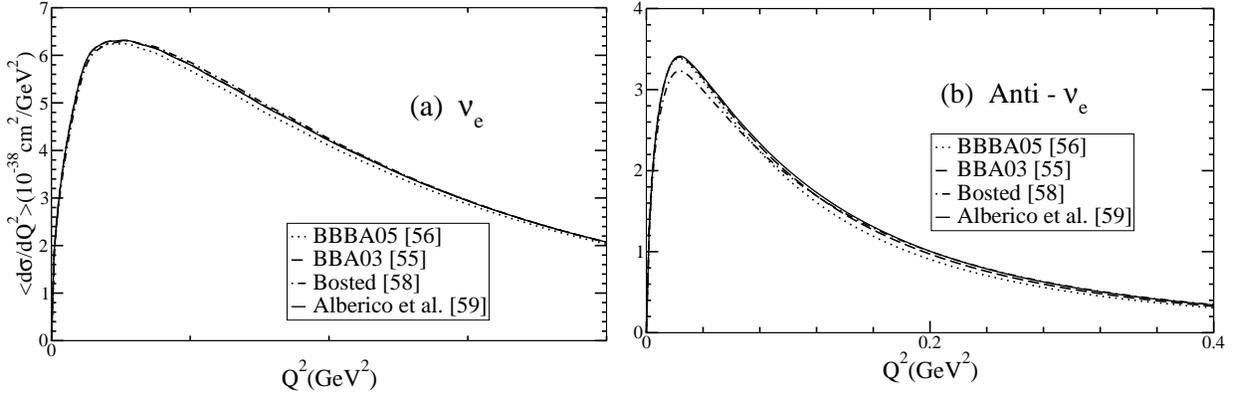}
\caption{$<\frac{d\sigma}{dQ^2}>$ vs $Q^2$ for the quasielastic process induced by electron type (a) neutrino and (b) antineutrino with various parameterization of the isovector form factors, with $M_V$=0.84GeV and $M_A$=1.1GeV.}
\label{fg:fig11}
\end{figure*}

We have shown in Figs.(\ref{fg:fig11}a) and (\ref{fg:fig11}b), the effect of various parameterizations of the isovector vector form factors on the $Q^2$ distribution in the case of charged current quasielastic lepton production process induced by $\nu_e$ and $\bar\nu_e$ in $^{16}$O calculated in the local Fermi gas model with RPA effects. The results have been shown with the parameterizations given by Budd et al.~\cite{BBA03}, Bradford et al.~\cite{BBBA05}, Bosted et al.~\cite{Bosted} and Alberico et al. \cite{Alberico}. 
We find that the use of various parametrization for the isovector form factor results in a very small change in the $Q^2$ distribution in the peak region. 

\begin{figure*}
\includegraphics{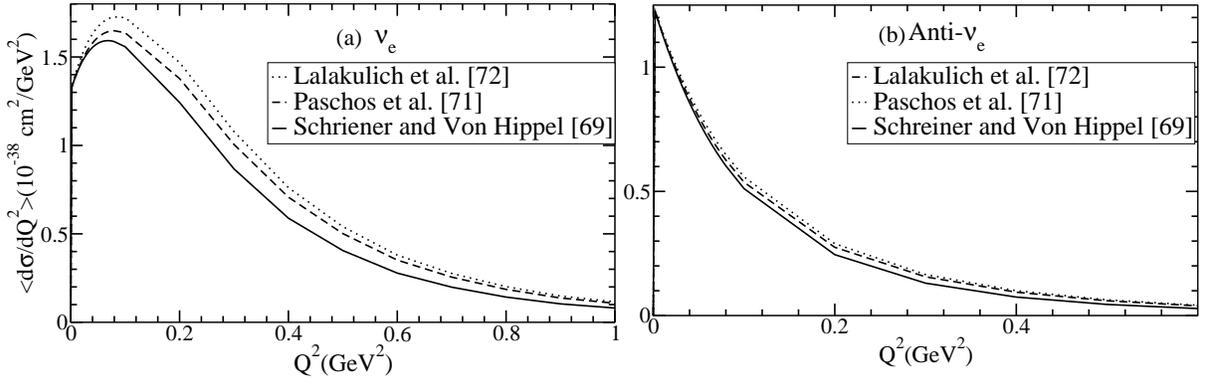}
\caption{$<\frac{d\sigma}{dQ^2}>$ vs $Q^2$ for incoherent process induced by electron type  (a) neutrino and (b) antineutrino averaged over the atmospheric neutrino flux with various form factors.}
\label{fg:fig12}
\end{figure*}

In Figs.(\ref{fg:fig12}a) and (\ref{fg:fig12}b), the dependence of the various parameterizations of the N-$\Delta$ transition form factors on the $Q^2$-distribution in the case of charged current 1$\pi^+$  production process induced by $\nu_e$ and 1$\pi^-$  production process induced by $\bar\nu_e$ in $^{16}$O with nuclear medium and pion absorption effects have been shown. These results are obtained by using the N-$\Delta$ transition form factors parameterizations given by Lalakulich et al.~\cite{Lalakulich}, Paschos et al.~\cite{Pasff} and Schreiner et al.~\cite{schvon}. We find that in the case of $\nu_e$ induced pion production process the differential cross section obtained by Paschos et al.~\cite{Pasff}  is 5-7$\%$ smaller in the region of $Q^2 \approx$0.1-0.5GeV$^2$, while the differential cross section obtained by Schreiner et al.~\cite{schvon} is 5$\%$ smaller at low Q$^2$ and which increases to around 10-16$\%$ for Q$^2$=0.2-0.4GeV$^2$, than the cross section calculated with Lalakulich et al.~\cite{Lalakulich} parameterization. In the case of $\bar\nu_e$ induced process the use of various parametrization for the N-$\Delta$ transition form factors results in a very small change in the peak region of $Q^2$ distribution. 

\begin{figure*}
\includegraphics{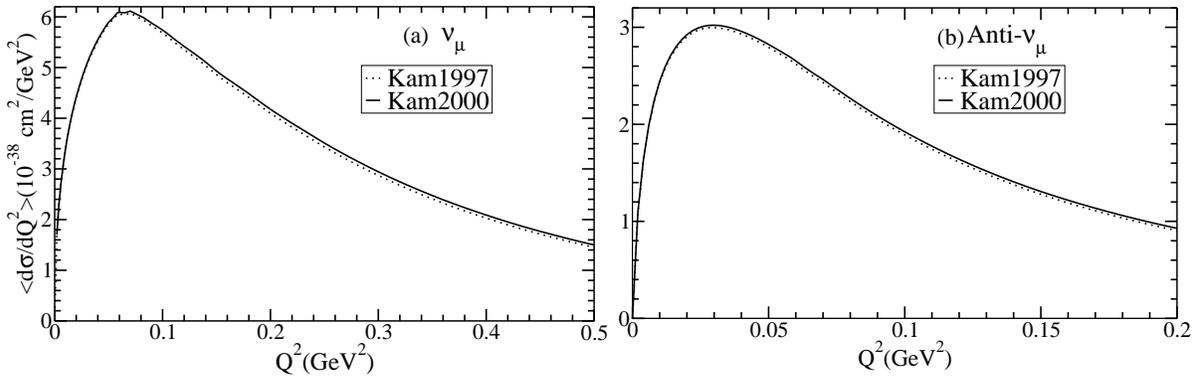}
\caption{$<\frac{d\sigma}{dQ^2}>$ vs $Q^2$ for quasielastic process induced by muon type (a) neutrino and (b) antineutrino averaged over the atmospheric neutrino flux for the solar minimum and solar maximum given by Honda et al.~\cite{Honda1}, \cite{Honda2}.}
\label{fg:fig13}
\end{figure*}

To show the dependence of the different fluxes at the Superkamiokande site for the Solar minimum and Solar maximum defined by Kam1997 and Kam2000 by Honda et al.~\cite{Honda1}, we have obtained the numerical results for the $Q^2$ distribution in the case of charged current quasielastic lepton production process induced by $\nu_\mu$ and $\bar\nu_\mu$ reactions in $^{16}$O calculated in the local Fermi gas model with RPA effects. The results for $\nu_\mu$ induced process is shown in Fig.(\ref{fg:fig13}a) and for the $\bar\nu_\mu$ induced process the results are shown in Fig.(\ref{fg:fig13}b). We find that these two fluxes result in a very small difference in the $Q^2$ distribution. Similarly in the case of inelastic one pion production cross section we find that the difference in the numerical results calculated using these two fluxes is very small.

\begin{table*}
\begin{tabular}{|c|c|c|c|c|}\hline\hline

& Muon Events & & &\\\hline \hline
&Process&Free Case&FGM&FGM With RPA\\ \hline

I&$\nu_{\mu}n \rightarrow \mu^{-}p(in ^{16}O)$&3332&2472&1894\\ \hline

II&$\bar \nu_{\mu}p \rightarrow \mu^{+}n(in ^{16}O)$&966&620&461\\\hline

III&$\bar \nu_{\mu}p \rightarrow \mu^{+}n$(on free p due to $H_{2}$)&241&241$^{\dagger}$&241$^{\dagger}$\\ \hline\hline
& $\nu_{\mu}+\bar \nu_{\mu}$&4539&3232&2596\\\hline\hline
& Electron Events & & &\\\hline \hline
&Process&Free Case&FGM&FGM With RPA\\ \hline
IV& $\nu_e n \rightarrow e^{-}p$(in $^{16}O$)&~~~2332& ~~~ 1754& ~~~ 1278\\\hline
V& $\bar \nu_e p \rightarrow e^{+}n$(in $^{16}O$) & ~~~609& ~~~ 358& ~~~ 266\\\hline
VI& $\bar \nu_e p \rightarrow e^{+}n$(on free p due to $H_2$)&~~~152& ~~~152$\dagger$& ~~~152$\dagger$\\
\hline\hline
&$\nu_e + \bar \nu_e$& ~~~3093& ~~~2264& ~~~1696\\
\hline\hline
\end{tabular}
\caption{Total number of lepton events for a quasielastic process. $\dagger$: For reaction on free protons the events would be the same in all the three columns.}
\end{table*}
\begin{table*}
\begin{tabular}{|c|c|c|c|}\hline
& Muon Events& &\\\hline\hline
&Inelastic Process&Free Case&Medium effects\\
 & & & with Pion abspn\\ \hline

VII&$\nu_{\mu}p \rightarrow \mu^{-}\Delta^{++}$(on free p due to $H_{2}$)&~~~154 &~~~154\\ \hline
VIII&$\bar \nu_{\mu}p \rightarrow \mu^{+}\Delta^{0}$(on free p due to$ H_{2}$)&~~~12 &~~~12\\\hline
IX&$\nu_{\mu} ^{16}O$($\mu^{-}$ accompanied by $\pi^{0}$) & ~~~171&~~~92\\ \hline
X&$\bar \nu_\mu ^{16}O$ ($\mu^{+}$ accompanied by $\pi^0$& ~~~40&~~~23\\\hline
XI&$\nu_\mu ^{16}O$($\mu^{-}$ accompanied by $\pi^+$) &~~~756 &~~~409\\\hline
XII&$\bar \nu_\mu ^{16}O$($\mu^{+}$ accompanied by $\pi^-$)&~~~179 &~~~102\\\hline
XIII&$\nu_\mu +\bar \nu_\mu$(Coherent) &~~~233 &~~~30\\\hline
XIV&$\nu_\mu +\bar \nu_\mu$ (Quasielastic like events from Inelastic Process) &~~~-& ~~~344\\
\hline\hline
&$\nu_\mu +\bar \nu_\mu$& ~~~1545& ~~~1166\\
\hline\hline
& Electron Events& &\\\hline\hline
& Inelastic Process& Free case & ~~~Medium Effects\\
&&& with Pion Abspn\\\hline
\hline
XV& $\nu_e p \rightarrow e^{-}  \Delta^{++}$(on free p due to $H_2$)&~~~98& ~~~98\\\hline
XVI& $\bar \nu_e p \rightarrow e^{+} \Delta^0$(on free p due to $H_2$) &~~~6& ~~~6\\\hline
XVII& $\nu_e ^{16}O$ ($e^{-}$ accompanied by $\pi^0$)&~~~99& ~~~53\\\hline
XVIII&$\bar \nu_e ^{16}O$ ($e^{+}$ accompanied by $\pi^0$)& ~~~21& ~~~ 12\\\hline
XIX& $\nu_e ^{16}O$ ($e^{-}$ accompanied by $\pi^+$)&~~~501& ~~~269\\\hline
XX&$\bar \nu_e ^{16}O$ ($e^{+}$ accompanied by $\pi^-$) &~~~91& ~~~52\\\hline
XXI&$\nu_e +\bar \nu_e$ (Coherent)& ~~~148& ~~~19\\\hline
XXII&$\nu_e +\bar \nu_e$ (Quasielastic like events from inelastic process)& ~~~-& ~~~200\\\hline
\hline
&$\nu_e +\bar \nu_e$& ~~~964& ~~~709\\ \hline
\hline
\end{tabular}
\caption{Total number of lepton events for inelastic process.}
\end{table*}
\begin{table*}
\begin{tabular}{|c|c|c|}\hline\hline
 Process& $\nu_{e} + \bar \nu_{e}$ &$\nu_{\mu} + \bar \nu_{\mu}$ \\ \hline\hline
Free case(QE+Inelastic)& ~~~4057& ~~~6084\\\hline
FGM without RPA+Inelastic with nuclear medium& ~~~ 2973& ~~~4499\\
and final state interaction effects& &  \\\hline
FGM with RPA +Inelastic with nuclear medium& ~~~2405& ~~~3762\\
and final state interaction effects& &  \\\hline
Monte Carlo events& ~~~2533.9& ~~~3979.7\\\hline
Reported by experiments& ~~~3353& ~~~3227\\
\hline\hline
\end{tabular}
\caption{Total number of lepton events calculated in our model and its comparison with the observed lepton events by SuperK collaboration and the Monte carlo number used by them~\cite{Atmos2}.}
\end{table*}
\subsection{Total lepton events}
Here we are going to present the results for the total number of lepton events for the sub-GeV energy region. These results have been presented for a 22.5kT water fiducial mass for 1489 days, and we have put a cut on the lepton's energy $E_l~<~1.33GeV$ and momenta of electrons and muons as $p_e>$100MeV and $p_\mu>$200MeV. We have integrated the total scattering cross section $\sigma$ over the atmospheric neutrino flux given by Honda et al.~\cite{Honda2} for the SuperK cite. 

The flux averaged cross section is defined as
\begin{equation}\label{avg_xsection}
<\sigma>=\int_{{E_\nu}_{min}}^{{E_\nu}_{max}} \sigma(E) \phi(E) dE
\end{equation}
where $\sigma(E)$ is the total scattering cross section and $\phi(E)$ is the atmospheric neutrino flux.

In Table-I the lepton event rates have been obtained for the CCQE processes induced by $\nu_{l}$ and $\bar \nu_{l}$ ($l=e,\mu$). In Table-II the lepton event rates have been obtained for the CC1$\pi$ production due to $\nu_{l}$ and $\bar \nu_{l}$ ($l=e,\mu$)induced reactions in $^{16}O$, as well as the leptons obtained from $\nu_{l}$ and $\bar \nu_{l}$ ($l=e,\mu$)induced reactions on the free protons and the leptons accompanied by $\pi^0$ in the neutrino(antineutrino) induced processes. In Table-III, we have presented the total lepton events from $\nu_{l}+\bar \nu_{l}$ (e or $\mu$) induced quasielastic and inelastic pion production processes. Here we have compared our final results (leptons obtained from CCQE reaction in local FGM with RPA effect + CC1$\pi$ production with nuclear medium and final state interaction effects) with the experimentally observed lepton events by SuperK collaboration and also with the lepton events used in the Monte Carlo analysis of these events by the SuperK collaboration~\cite{Atmos2}. 

In our calculations for predicting the lepton events, we have considered the following channels. In the quasielastic process the contributions to the lepton events have been taken from the channels ~(i) ~$\nu_l n \rightarrow l^- p(in~ ^{16}O)$, ~(ii)~ $\bar \nu_l p \rightarrow l^+ n(in~ ^{16}O)~$ and ~(iii)~$\bar\nu_l p \rightarrow l^{+} n~$(on free p due to ~$H_{2}$~). In the case of incoherent pion production process the various channels contributing to the lepton events are ~(i) ~$\nu_l p \rightarrow l^{-} \Delta^{++}~$(on free p due to ~$H_{2}~$), ~(ii)~ $\bar \nu_l p \rightarrow l^{+} \Delta^{0}~$(on free p due to~$ H_{2}~$), (iii)~$\nu_l~ ^{16}O$ ($l^-$ accompanied by $\pi^{0}$~), (iv) ~$\bar\nu_l ~^{16}O$ ($l^+$ accompanied by $\pi^0$~), ~(v)~$\nu_l ~^{16}O$ (~$l^-$ accompanied by $\pi^+$~), ~(vi)~$\bar \nu_l~ ^{16}O$ (~$l^+$ accompanied by $\pi^-$). In the case of coherent process the contributions would come from  ~(i) ~$\nu_l + ^{16}O \rightarrow l^- + \pi^+ +^{16}O$ and ~(ii) ~$\bar\nu_l + ^{16}O \rightarrow l^+ + \pi^- +^{16}O$ processes.

In the case of incoherent and coherent pion production processes, when pion absorption effects are taken into account, the pions which are produced but get absorbed while coming out of the nucleus, the reaction produces only lepton. Such reactions give leptons without the pions in the final state and are labelled as quasielastic like events. We have also considered quasielastic like events in calculating the event rates. The effect of nuclear medium effects on the total number of lepton events can be summarised by saying that it leads to a reduction of 40$\%$ in the event rate.
\section{Conclusions}
In this work we have studied the nuclear effects in the charged lepton production in water induced by atmospheric neutrinos at the SuperK site. The energy dependence for the total cross section for the quasielastic, incoherent and coherent pion productions have been calculated in the local density approximation for neutrino(antineutrino) induced reactions in $^{16}O$. We find that nuclear medium effects play a very important role in the study of $\sigma$ as well as flux averaged differential cross section $<\frac{d\sigma}{dQ^2}>$. The total lepton production event rate is compared with the experimental observed numbers at SuperK and also with the numbers used in their Monte Carlo\cite{Atmos2}.

We conclude the following:

(i) In the case of neutrino induced charged current quasielastic lepton production process, the nuclear medium effects like Pauli blocking, Fermi motion effects, renormalization of weak transition strengths in the nuclear medium, reduces the cross section. This reduction is large in the energy region of $E_\nu$= 0.4-0.5GeV than the reduction in the energy region of $E_\nu$=1-3GeV as compared to the cross section calculated for the free case.  However, in the case of antineutrino this reduction is more than in the case of neutrino as compared to the cross section calculated for the free case. 
In the incoherent charged current lepton production process accompanied by a pion, the reduction in the cross section due to the nuclear medium effects is more than the reduction due to the final state interaction of pions. While in the case of coherent pion production, the reduction due to final state interaction is quite large as compared to the reduction due to the nuclear medium effects. Furthermore, we find that the contribution of the cross section calculated in the case of coherent pion production process in the nuclear medium and final state interaction effects is around 6-7$\%$ to the total (incoherent+coherent) one pion production cross section in the energy region of $0.4GeV<E<3GeV$, and due to this we have not considered the form factor dependence, $Q^2$ distribution, etc. in this case.
 
(ii) We find that in the case of neutrino, the Q$^2$ distribution calculated in the local Fermi gas model with RPA effects results in a large reduction in the peak region of $Q^2$. For antineutrino induced process, there is a shift in the peak region which is towards low Q$^2$ and the reduction is smaller than in the case of neutrino induced process. In the case of incoherent one pion production process, there is a reduction in the peak region of $Q^2$ by taking medium effects into account which becomes smaller at large value of Q$^2$. When pion absorption effects are also taken along with medium effects then there is a further reduction in the distribution. For antineutrino induced process, peak shifted towards low Q$^2$ and the reduction is smaller than in the case of neutrino induced process.   

(iii) The dependence of the axial dipole mass $M_A$ on the total scattering cross section for the neutrino induced charged current quasielastic process and for the incoherent charged current lepton production process in $^{16}O$ is studied. We find that the scattering cross section increases when $M_A$ is taken as 1.21GeV, while it decreases when $M_A$=1.05GeV as compared to the cross sections calculated with $M_A$=1.1GeV. In the case of antineutrino induced reaction the dependence on $M_A$ becomes small. 

(iv) The dependence of the various parameterizations of the isovector form factors on the $Q^2$ distribution in the case of charged current quasielastic lepton production process and N-$\Delta$ transition form factors in the case of charged current 1$\pi^+$(1$\pi^-$) production process induced by neutrino(antineutrino) in $^{16}$O has been studied. For the quasielastic process results have been presented with the parameterizations given by Budd et al.~\cite{BBA03}, Bradford et al.~\cite{BBBA05}, Bosted et al.~\cite{Bosted} and Alberico et al. \cite{Alberico}. We find that the use of various parametrization for the isovector form factor in the case of charged current quasielastic lepton production process results in a very small change in the peak region of $Q^2$. 

In the case of incoherent one pion production process these results are obtained by using the N-$\Delta$ transition form factor parameterizations given by Lalakulich et al.~\cite{Lalakulich}, Paschos et al.~\cite{Pasff} and Schreiner et al.~\cite{schvon}. We find that in the case of neutrino induced pion production process the differential cross section obtained by Paschos et al.~\cite{Pasff} and Schreiner et al.~\cite{schvon} is smaller in the region of low Q$^2$, which increases by small amount for Q$^2$=0.2-0.4GeV$^2$, than the cross section calculated with Lalakulich et al.~\cite{Lalakulich} parameterization. In the case of antineutrino induced process there is very small change in the $Q^2$ distribution.

(v) Nuclear medium effects play an important role in reducing the number of events obtained by integrating $\sigma$ over the atmospheric neutrino flux. In the case of $\nu+\bar\nu$ induced quasielastic lepton production the reduction is around $25\%$ when the events are calculated in the local Fermi gas model without the RPA effects and a total reduction of around $45\%$ when RPA effects are also taken into account in comparison to the lepton events calculated for the free case. In the case of inelastic lepton production the reduction from the events calculated without medium effects is around $40\%$ in the case of incoherent pion production and around 85$\%$ in the case of coherent pion production, which results in a net reduction of around 50$\%$ when nuclear medium and final state interaction effects are taken into account. The lepton events from the inelastic process also contribute to the quasielastic events even when there are no pions in the final state because of its absorption while coming out of the nucleus. Such quasielastic like events contribute around 12$\%$ to the total quasielastic lepton events.

We find that for the total lepton events obtained by integrating the total cross section cross section($\sigma$) over the atmospheric neutrino flux results in a large reduction in the event rates when RPA effects are taken into account. 
Therefore, the results with nuclear medium effects for muon and electron events may be important in the analysis of neutrino oscillation experiments. Thus, we conclude that in the future neutrino oscillation experiments to be performed at SuperK on the atmospheric neutrino or the accelerator neutrinos like T2K and NO$\nu$A, the study of nuclear medium effects in predicting the event rates would play a very important role. 

\subsection{Acknowledgments}
One of the authors(M. S. A.) is thankful to T. Kajita and Y. Hayato (I. C. R. R., University of Tokyo) for many useful discussions and the warm hospitality provided during his stay at ICRR where part of this work was done. Thanks are also due to M. Honda for providing us atmospheric neutrino flux and many useful discussions. S. C. is thankful to the Jawaharlal Nehru Memorial Fund for the Doctoral Fellowship.

\end{document}